\documentclass[a4paper,12pt]{article}
\usepackage{bbm} 
\usepackage[utf8]{inputenc}
\usepackage{amssymb,amsmath,amsfonts}
\usepackage[colorlinks,linkcolor=blue,citecolor=Green,urlcolor=blue]{hyperref}
\usepackage{color}
\usepackage{graphicx}
\usepackage{hyphenat}
\usepackage[dvipsnames]{xcolor}
\usepackage{wrapfig}
\usepackage{empheq}
\usepackage{textcomp}
\usepackage[caption=false]{subfig}
\usepackage{rotating}
\usepackage{wrapfig}
\usepackage{pdfpages}
\usepackage{verbatim}
\usepackage{setspace}
\usepackage{array}
\usepackage{caption}
\usepackage[pdf]{pstricks}
\usepackage{upgreek}
\usepackage{cmll}
\usepackage{latexsym}
\usepackage{braket}
\usepackage{cite}
\usepackage{epsfig}
\usepackage[left=2.5cm,top=3cm,right=2.5cm,bottom=3cm,bindingoffset=0cm]{geometry}
\usepackage[titletoc,page]{appendix}
\usepackage{multicol}
\usepackage{youngtab}

\newcommand{\beq}{\begin{eqnarray}}
\newcommand{\eeq}{\end{eqnarray}}
\newcommand{\bea}{\begin{eqnarray}}
\newcommand{\eea}{\end{eqnarray}}
\newcommand{\be}{\begin{equation}}
\newcommand{\ee}{\end{equation}}

\def\brc{\langle}
\def\ckt{\rangle}
\def\const{{\rm const}}
\def\de{\partial}

\def\Tr{\qopname\relax o{Tr}}

\numberwithin{equation}{section}

\numberwithin{equation}{section}

\begin{document}

\title{\bf  Large Angular Momentum}  

\author{Kenichi Konishi$^{(1,2)}$, Roberto Menta$^{(3,4)}$ 
  \\[13pt]
{\em \footnotesize
$^{(1)}$Department of Physics ``E. Fermi", University of Pisa, Largo Pontecorvo, 3, Ed. C, 56127 Pisa, Italy}\\[2pt]
 {\em \footnotesize
$^{(2)}$INFN, Sezione di Pisa, Largo Pontecorvo, 3, Ed. C, 56127 Pisa, Italy}\\[2pt]
 {\em \footnotesize
$^{(3)}$ Scuola Normale Superiore, Piazza dei Cavalieri, 7, 56127 Pisa, Italy}\\[2pt]
 {\em \footnotesize
$^{(4)}$Laboratorio NEST,  Piazza San Silvestro, 12, 56127 Pisa, Italy}\\[2pt]
\\[-5pt]
\\[1pt] 
{  
\footnotesize  \texttt{kenichi.konishi@unipi.it}, \ \  \texttt{roberto.menta@sns.it}
 }  
}
\date{}

\maketitle

\begin{abstract}
The quantum states of a spin $\tfrac{1}{2}$ (a qubit) are parametrized by the space ${\mathbf {CP}}^1 \sim  S^2$,  the Bloch sphere. A spin $j$ for a generic $j$  (a $2j+1$-state system)  is represented  instead by a point of a larger space,  ${\mathbf {CP}}^{2j}$.  Here we study the state of a single angular momentum/spin in the limit,  $j \to \infty$.  The special class of states  $ | j, {\mathbf  n}\ckt   \in  {\mathbf {CP}}^{2j} $, with spin oriented towards 
definite spatial directions  ${\mathbf n} \in  S^2$,
i.e.,    $({\hat  {\mathbf  J}}\cdot {\mathbf  n} ) \,  | j, {\mathbf  n}\ckt   = j\,  |j,  {\mathbf  n}\ckt $,  are found to behave as  classical angular momenta,  $j \, {\mathbf n}$, in this limit.   Vice versa,  general spin states in  ${\mathbf {CP}}^{2j}$ do not become classical, even at large $j$.
We study these questions, by analysing the Stern-Gerlach processes, the angular-momentum composition rule,  and the rotation matrix.  
 Our observations help to clarify better how classical mechanics emerges from quantum mechanics in this context  (e.g., with unique trajectories of a particle carrying a large spin in an inhomogeneous magnetic field), and to make  the widespread idea that large spins somehow become classical,  more precise.
\end{abstract}

\newpage  

\tableofcontents

\newpage

\section{Introduction}  

It is a widely shared view that in the limit of large spin (angular momentum) a quantum-mechanical spin somehow becomes classical
 \cite{Brussard}-\cite{Corso}.
After all, the large spin limit ($j/\hbar \to \infty$)   is equivalent to the $\hbar \to 0$ ($j$ \,fixed)  limit, and, in accordance with the general idea of  the semi-classical limit in quantum mechanics, a large-angular-momentum state might be  expected to behave as a classical  angular momentum. But to the best of the authors' knowledge, it has never been clearly shown  whether  this indeed occurs,  or if it does, exactly how.    It is the aim of this note to help filling in this gap.

The question was put under a sharp focus in a general  discussion on emergence of classical mechanics for a macroscopic body from quantum mechanics,  
as formulated in  \cite{KK2,KKHTE}.  Knowing about the typical Stern-Gerlach (SG) process for a spin $\frac{1}{2}$ atom,  one asks what happens if a macroscopic body consisting of many (e.g.~$N \sim \mathcal{O}(10^{23}))$ spins is set under an inhomogeneous magnetic field of a SG set-up. 
A classical particle  with a magnetic moment directed  towards    
\be     
{\boldsymbol {\mu}} = \const. \,   {\bf n}\;, \qquad  {\bf n} =  (\sin \theta  \cos \phi,  \sin \theta  \sin \phi,   \cos \theta)\;, 
\ee
moves  in an inhomogeneous magnetic field
according to Newton's equations, 
\be   
m\, {\dot {\bf r}} = {\bf p}\;, \qquad    \frac{d {\bf p}}{dt} =  {\bf  F}  =  \nabla ( {\boldsymbol {\mu}}  \cdot {\bf B} )\;. \label{Newton} 
\ee
The way such  a unique trajectory for a classical particle emerges from quantum mechanics has been discussed in Ref.~\cite{KK2}. The magnetic moment of a macroscopic particle is the expectation value 
\be         
 \sum_i   \langle \Psi |    ( {\hat   {\boldsymbol \mu}}_i +     \frac{e_i    {\hat  {\boldsymbol \ell}}_i  }{2  m_i  c } ) |\Psi \rangle  =   {\boldsymbol  \mu}\;,    \label{classicalSG}
\ee
taken in the internal bound-state wave function $| \Psi \rangle$ where $\mu_i$ and  $\frac{e_i    {\bf \ell}_i  }{2  m_i  c }$ denote the intrinsic magnetic moment and the one due to the orbital motion of the $i$-th constituent atom or molecule ($i=1,2,\ldots, N$), respectively. Clearly, in general, the well-known picture for a spin $\frac{1}{2}$ atom, with a doubly split wavepacket, does not generalize to a classical body (\ref{classicalSG}) with  $N \sim \mathcal{O}(10^{23})$ constituents. 

Typically, the spins of the component atoms and molecules are oriented along different directions, but the crucial point is that they are bound in the macroscopic body  in various atomic, molecular or crystalline structures. The wave function of particles in bound states does not  split in sub-wavepackets,  in contrast to that of an isolated  spin $\tfrac{1}{2}$ atom, under an inhomogeneous magnetic field. The bound-state Hamiltonian does not allow it \footnote{This discussion is  closely parallel to the question of quantum diffusion.
Unlike free particles, particles in bound states (the electrons in atoms;  atoms in molecules, etc.) do not diffuse, as they move in binding potentials. This is one of the elements for the emergence of the classical mechanics, with unique trajectories  for macroscopic bodies. As for the center-of-mass (CM) wave function of an isolated macroscopic body, its free quantum diffusion is simply suppressed by  mass.}. What determines the motion of the CM of the body is the average magnetic moment, (\ref{classicalSG}). 

\begin{figure}[t]
\begin{center}
\includegraphics[width=4in]{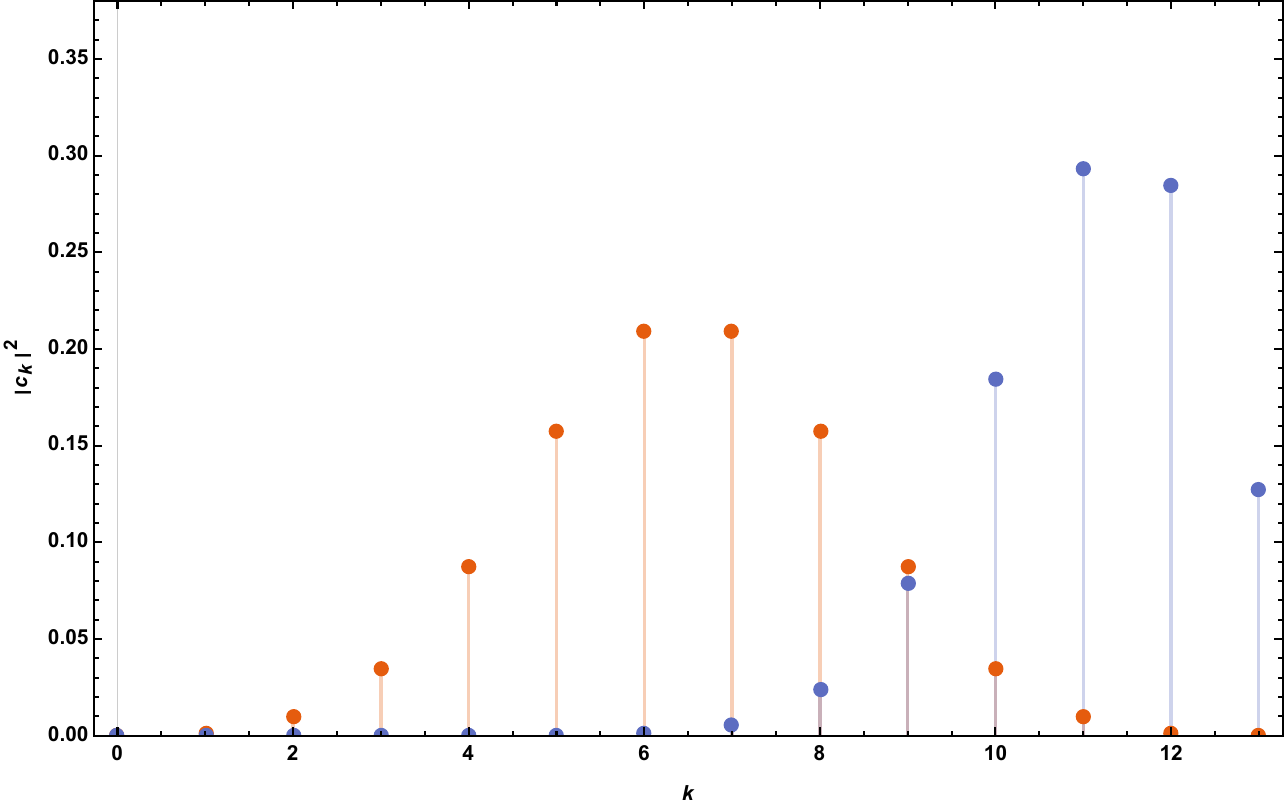}
\caption{\footnotesize  The distribution in $J_z= -j +k$  ($k=0,1,\ldots,  2j$)    for a spin $j=13/2$ particle in the state  (\ref{see00}), (\ref{exercise00}), with $\theta = \pi/2$ (center, orange dots) or with $\theta = \pi/4$  (right, blue dots). 
 The wave packet of such a particle spreads  into many  subwavepackets  \`a la Stern-Gerlach under an inhomogeneous magnetic field with strong gradient in the ${\hat z}$ direction.
}
\label{Spreadsi}
\end{center}
\end{figure}

Logically,  however, one cannot exclude special cases  (e.g. a magnetized piece of metal)  with all spins inside the body oriented in the same direction. One might wonder how  the wave function of  such a particle with spin $j  \to \infty$ behaves, under an inhomogeneous magnetic field.   
The  question is  whether the three conditions recognized in \cite{KK2} for the emergence of classical mechanics  for a macroscopic body  with a unique trajectory,  reported here  in  Appendix~\ref{NewtonEq},   are indeed sufficient.  Or,  is some extra condition,  or perhaps a new unknown mechanism,  needed,  to suppress possible wide spreading  of the wave function into many sub-packets  (such as in  Fig.~\ref{Spreadsi} for a small spin, $j=13/2$) under an inhomogeneous  magnetic field?

This is the central question we are going to investigate.  We shall  first discuss in Sec.~\ref{spaces} some mathematical and physical aspects of the quantum states of a small and large angular momenta, and compare them with the properties of a classical angular momentum.  Considerations of quantum fluctuations of various angular momentum states lead us to propose  that a particular class of states,
$ |j, {\mathbf  n}\ckt$, 
 \be   {\hat  {\mathbf  J}}^2   \,  | j, {\mathbf  n}\ckt  =  j(j+1)   | j, {\mathbf  n}\ckt \,,\qquad  
 ({\hat  {\mathbf  J}}\cdot {\mathbf  n} ) \,  | j, {\mathbf  n}\ckt   = j\,  |j,  {\mathbf  n}\ckt \,,      \label{Bloch}  \ee   
where ${\mathbf n}$ is a unit vector directed towards  $(\theta, \phi)$ direction, 
\be    {\mathbf n} =     (\sin \theta \cos \phi,  \sin \theta \sin \phi, \cos \theta)\;, 
\ee
or close to them,  are to be identified with a classical angular momentum, $j\, {\mathbf n}$, in the large $j$ limit. 

Vice versa,  generic states of large spin  will be found to remain quantum mechanical, with large fluctuations,  
even in the limit, $j\to \infty$, showing that the often stated belief that a large spin (angular momentum)
 becomes classical, should not be taken for granted, literally.  
 
 Lest the reader is led to a misunderstanding of the content of our work, let us make the following point clear. 
 The idea that a large spin is made of many spin $\tfrac{1}{2}$ particles is quite a fruitful one both from mathematical and physical point of view. From the mathematical point of view,  the entire theory of angular momentum can indeed be reconstructed this way  \cite{Schwinger}, and it will help to recover certain formulas for large angular momentum states easily.  It is also useful from physics point of view,  as such a system may be regarded as an idealized, toy  model for a more realistic macroscopic body, made of many atoms and molecules (carrying spins and orbital angular momenta).  And this picture helps to interprete some of our findings.
(The states of type (\ref{Bloch}) are known,  in such a context of many-body systems, as spin coherent states, Bloch states or  Glauber coherent states,  depending on the author  \cite{Radcliffe,Puri, Arecchi,Aravind,Lieb,Wodk,Livine,LohKim,Byrnes}.  They are all equivalent to the state of definite spin orientation (\ref{Bloch}), as far as the global spin property is concerned.)

Nevertheless,  our discussion is, as will be clear from an attentive reading, entirely about the quantum or classical properties of  {\it a single large spin (or angular momentum).}  We are not concerned here with the thermodynamical or other physical properties of a many-body system.
The main question is whether a single large spin system behaves classically or remains quantum mechanical, in the limit, $j/\hbar \to \infty$.  As will be  seen in the following, the answer turns out to be quite subtle and nontrivial. 

In subsequent sections,    we are going to examine  these questions   
 through the analysis of the Stern-Gerlach processes  (Sec.~\ref{SG}), the orbital angular momenta  (Sec.~\ref{Orbital}),  the angular-momentum addition rule  (Sec.~\ref{Addition}),  and the rotation matrix  (Sec.~\ref{Rotation}).  
Generic spin  $j$ states  far from (\ref{Bloch}) are shown to remain quantum mechanical even in the limit $j\to \infty$: the fate of these states will be discussed in 
 Sec.~\ref{GenericLS}.  Conclusion and a few more general reflections are in Sec.~\ref{Conclusion}.

\section{Space of quantum spin states and classical angular momenta \label{spaces}}

 The generic pure  spin $\frac{1}{2}$ (two-state system) state is described by 
 {\color{blue}  the vector} 
\be     
|\psi\ckt  =    c_1 |\!\uparrow\ckt +  c_2 |\!\downarrow\ckt \;, \label{pure1/2} 
\ee
where 
\be  (c_1, c_2) \sim  \lambda  (c_1, c_2)\;, \qquad   \lambda \in {\mathbf C}{\backslash}{\{0\}}\, \ee
and $|\!\uparrow\ckt $ and   $|\!\downarrow\ckt$  are  spin up and spin down states, i.e., the eigenstates of ${\hat J}_z$ with eigenvalues, $\pm \frac{1}{2}$. 
 The complex numbers  $(c_1, c_2)$   describe the homogeneous coordinates  of the space  $\mathbf{CP}^1$ ($\sim S^2$).   We recall  that  any  state  (\ref{pure1/2}) can be interpreted as the state in which the spin is oriented in some direction ${\mathbf n}$, 
 \be   {\mathbf n}=(\sin \theta \cos \phi, \sin \theta \sin \phi, \cos \theta)\;,   \label{stella0} 
 \ee
 that is
 \be  
 {\hat J}^2   \left|\frac{1}{2}, {\mathbf  n}\right\ckt   =  \frac{3}{4}   \left|\frac{1}{2}, {\mathbf  n}\right\ckt \;, \qquad  \left({\hat  {\mathbf  J}}\cdot {\mathbf  n} \right) \,  \left|\frac{1}{2}, {\mathbf  n}\right\ckt   =   \frac{1}{2} \,  \left|\frac{1}{2}, {\mathbf  n}\right\ckt   \;,\label{corresp2}
\ee
 \be      
|{\bf n}\ckt    \equiv     \left|\frac{1}{2}, {\mathbf  n}\right\ckt   =    c_1 |\!\uparrow\ckt +  c_2 |\!\downarrow\ckt \;,   \qquad  c_1=   e^{-i \phi/2}  \cos \tfrac{\theta}{2}\;, \qquad c_2=   e^{i \phi/2}  \sin \tfrac{\theta}{2}\;.   \label{WFspin120}  
\ee
 without loss of generality.  This is so  because the  space of pure spin  $\frac{1}{2}$ quantum states and that of the unit space vector coincide: they are both $S^2 \sim \mathbf{CP}^1$. Indeed, given the state (\ref{WFspin120}), there is always  a rotation matrix such that 
 \be    
D^{1/2}(0, \theta,\phi)  \cdot   \left(\begin{array}{c} e^{-i \phi/2}  \cos \tfrac{\theta}{2}\ \\ e^{i \phi/2}  \sin \tfrac{\theta}{2}\end{array}\right)  =    \left(\begin{array}{c}1 \\0\end{array}\right)\;,   \label{leads}
 \ee
 where the general rotation matrix for a spin $\frac{1}{2}$  is given, using Euler angles \footnote{Note that the third Euler angle  (the rotation angle $\gamma$  about the final $z$-axis) is redundant here, as it gives only the phase $e^{i\gamma/2}$, and has been set to $0$ in Eq.(\ref{leads}).} ($\gamma, \theta, \phi$),  by
\be 
D^{1/2}(\gamma, \theta,\phi)    =  U_z(\gamma)   U_y(\theta) U_z(\phi)=  \left(\begin{array}{cc}e^{i (\gamma + \phi)/2} \cos \tfrac{\theta}{2} & e^{(\gamma - \phi)/2}  \sin \tfrac{\theta}{2}  \\- e^{i(-\gamma+\phi)/2} \sin\tfrac{\theta}{2} & e^{-i (\gamma+\phi)/2} \cos \tfrac{\theta}{2}\end{array}\right)\;.   \label{rotationM}  \ee  
The fact that any spin $\frac{1}{2}$ quantum state can be associated with a definite space direction, however, does not mean that it can be identified with a classical angular momentum,  $\frac{1}{2} \hbar  {\mathbf n}$, as the fluctuations in the  directions transverse to ${\mathbf n}$  are always of the same order of the spin magnitude itself (see Eq.(\ref{strong}) below, with $j=\frac{1}{2}$, $m =\pm  \frac{1}{2}$).   It is always a quantum mechanical system.

Let us now consider a generic spin $j$ state.
It is a special type of $2j+1$-state 
quantum system.  Its general wave function has the form, 
\be 
 |\psi \ckt   =   \sum_{m=-j}^{j}   b_m    |j, j_z=m \ckt   \label{generalj} 
\ee
where 
\be   
(b_{-j},b_{-j+1}, \ldots, b_j) \sim   \lambda  (b_{-j},b_{-j+1}, \ldots, b_j) \;, \qquad     \lambda \in  {\mathbf C}{\backslash}{\{0\}} \;,    \label{cpn}
\ee
are the coordinates of the complex projective space $\mathbf{CP}^{2j}$ \footnote{  
As is well known,  the manifold of pure quantum states of any $n$-state system is $\mathbf{CP}^{n-1}$  \cite{Bengt}.   The spin states (\ref{cpn})   are special cases
with $n=2j+1$.  
  }.
On the other hand,  the variety of the directions in the three-dimensional space, ${\mathbf n}$, is always $S^2\sim \mathbf{CP}^1$.  It means that   this time 
not all states of spin $j$ state, (\ref{generalj}),   can be transformed by a rotation matrix (selecting an appropriate new ${\hat z}$ axis) into the form: 
\be   (b_{-j}^{\prime},b_{-j+1}^{\prime}, \ldots, b_j^{\prime}) \sim 
(0,0,\ldots, 0, 1) \;.   \label{like} 
\ee
This observation is the first hint  that  there are some  subtleties  in the way classical picture of angular momentum emerges 
from quantum mechanics at large $j$. It is the purpose of the present note to
elaborate on this point.

Let us start with the basic properties of quantum mechanical angular momentum.  These are of course well-known textbook materials.   Because of the commutation relations
\be  
 [{\hat J}_k, {\hat J}_{\ell}]=  i \hbar  \, \epsilon^{k \ell m}    {\hat J}_{m}   \;,  \label{commutation} 
\ee
only one of the components, e.g. ${\hat J}_z$, can be diagonalized together with the total (Casimir) angular momentum, ${\hat J}^2=  {\hat J}_x^2 +  {\hat J}_y^2 +  {\hat J}_z^2$, 
\be    
{\hat J}^2   |j,m\ckt = j(j+1) |j,m\ckt\;, \qquad   {\hat J}_z  |j,m\ckt = m   |j,m\ckt \;.   \label{generic} 
\ee
In the state,  $|j,m\ckt$, where  ${\hat J}_z$ has a definite value,  $m\, \hbar$,  ${\hat J}_x$ and  ${\hat J}_y$  are fluctuating.  Their magnitudes 
$ \Delta   {J}_{x,y} =       \brc j, m  |  {\hat J}_{x,y}^2 | j,m \ckt ^{1/2}$  are given by  
\be   
 \Delta   {J}_{x}  =   \Delta   {J}_{y}  =  \left(  \brc j, m  | \big({\hat J}^2 -  {\hat J}_z^2\big)   | j,m \ckt / 2 \right)^{1/2}   = \left(j(j+1) -  m^2)/2\right)^{1/2} \hbar    \sim  \mathcal{O}(j)   \hbar \;.  \label{strong}   
\ee
Namely, in a generic state $|j,m\rangle$, there are strong quantum fluctuations of $J_x$ and $J_y$ at $j \to \infty$, of the same order of magnitude with  $j$ itself.   There is no way such a state can be associated with a classical angular momentum, which requires all three components to be well-defined simultaneously.
The exception occurs for states $|j, \pm j \rangle$, for which
\be      
\Delta  {\hat J}_x = \Delta  {\hat J}_y =     (j/2)^{1/2}   \hbar    \ll j \, \hbar  \;.   \label{minimum} 
\ee 
For these state, it makes sense to interpret them as a classical angular momentum pointed towards $\pm {\hat z}$, as its transverse fluctuations $(\sim \mathcal{O}(\sqrt{j})$) are negligible with respect to its magnitude $j$, in the limit $j \to \infty$.   The same can be said of the states of spin almost oriented towards $\hat z$,  
$|j, \pm ( j - r)  \rangle$, $ r \ll  j$.

Naturally,  all states  of the form, $|j, {\mathbf  n}\ckt $,  in which spin is oriented towards a generic direction ${\mathbf n}$, 
\be  
 {\hat J}^2   |j, {\mathbf  n}\ckt   = j(j+1)   |j, {\mathbf  n}\ckt \;, \qquad  \left({\hat  {\mathbf  J}}\cdot {\mathbf  n} \right) \,  |j, {\mathbf  n}\ckt   = j\,  |j, {\mathbf  n}\ckt   \;,   
 \label{Bloch2}
\ee
  {\it   (already defined in  (\ref{Bloch})  in the Introduction)} 
have this property:  its fluctuations in the components perpendicular to  ${\mathbf n}$ become negligible in the  $j \to \infty$ limit.   The states (\ref{Bloch2}) are known as  
Bloch states or coherent spin states in the literature \cite{Radcliffe, Puri, Arecchi, Aravind, Lieb}.

These observations lead us to propose that  the quantum-classical  correspondence (by setting $\hbar=1$) to be made is 
\be       
  |j, {\mathbf n} \ckt   \quad   \longleftrightarrow     \quad    {\bf j}^{(\rm class)}=   j \, {\bf n}\;. 
\label{proposal}
\ee
Of course, all states ``almost oriented towards ${\mathbf n}$", $  |j, {\mathbf  n}, q  \ckt,$    
\be   
\left({\hat  {\mathbf  J}}\cdot {\mathbf  n} \right) \,  |j, {\mathbf  n}, q\ckt   = (j - q) \,  |j, {\mathbf  n}, q \ckt\;, \qquad  q=1,2,\ldots\,, \, \quad q \ll j  \label{corresp3}
\ee  
discussed in Sec.~\ref{almostOr} (see Eq.(\ref{almost})) below,  can also be interpreted as $ {\bf j}^{(\rm class)}=   j \, {\bf n}$.  We are going to examine below  the consistency of such a picture,  with respect to the SG processes  (Sec.~\ref{SG} below),  
a study of orbital  angular momentum (Sec.~\ref{Orbital}),  the addition rule of the angular momenta  (Sec.~\ref{Addition}) and the rotation matrix (Sec.~\ref{Rotation}).

A powerful idea, useful both for mathematical and physical reasoning below,  is that  the general spin $j$  can be regarded as a direct product state of  $n= 2j$ spins, each carrying $\frac{\hbar }{2}$.  From a group-theoretic point of view, it is well known that any irreducible representation of the SU(2) group,  which is the double covering group  (see e.g.,  \cite{Dubrovin}) of the  rotation
group SO(3),  is associated with the Young tableaux, 
\be   \underbrace{\yng(3)\cdots\yng(1)}_{n \ \text{boxes}} \;, \label{young}  
\ee 
with a single row of  $n$  boxes, i.e.,   the totally symmetric direct product of $n= 2j$  spin $\frac{1}{2}$ objects. 
Indeed,  the entire theory of quantum angular momentum can be reconstructed based on this point of view \cite{Schwinger},  
by using the operators, 
\be   
{\hat  J}^a  =    \sum_{i,j =1,2}    a_i^{\dagger}   \Big(\frac{\tau^a}{2}\Big)_{ij}   a_j \;,  \qquad  [{\hat  J}^a, {\hat  J}^b]= i\, \epsilon_{abc} {\hat J}^c\;, 
\label{analogous}  \ee 
where $\tau^a$ are the Pauli matrices, $a_i, a^{\dagger}_i$ are the harmonic-oscillator annihilation and creation operators, of spin-up ($i=1$) and spin-down ($i=2$) particles,
\be   
[a_i, a^{\dagger}_j]= \delta_{ij}\;.
\ee 
By construction,  all states $\propto    a_1^{\dagger}  a_2^{\dagger}    a_1^{\dagger}    a_1^{\dagger}  \ldots   |0\ckt$, with $k$ $a_1^{\dagger}$'s  and   $(N-k)$ $a_2^{\dagger}$'s  ($k=0,1,2,\ldots N$) are symmetric under exchanges of the spins, therefore belong to the irreducible representation ${\underline j}$, with $j= N/2$ \footnote{An analogous construction for the $SU(N)$ group, $N \ge 3$, is possible, but  only for totally symmetric representations, $\yng(3)\cdots\yng(1)$.  The usefulness of such an approach is somewhat limited compared to the $SU(2)$ case, where any irreducible representation has this form.  But  the construction analogous to (\ref{analogous}) for $SU(N)$  can explain, for instance,  the degeneracy of the energy eigenstates of  the $N$-
dimensional isotropic harmonic oscillator. }.

Physically, on the other hand,  one may consider (\ref{young})  as a particular state of  a toy-model macroscopic body made of $n$ atoms, each carrying spin $\frac{1}{2}$.  

Now a particularly attractive  picture of the quantum states (\ref{young}) of general spin $j$, follows from the so-called stellar representation of $\mathbf{CP}^{n}$   ($n=2j$)  \cite{Bengt}.  Let the complex numbers  
\be    (Z_0, Z_1, \ldots, Z_n)  \sim  \lambda \,   (Z_0, Z_1, \ldots, Z_n) \;,  \qquad    \lambda \in  {\mathbf C}{\backslash}{\{0\}} 
\label{rescal}     \ee   
be the homogeneous coordinates of $\mathbf{CP}^{n}$, and  consider the $n$-th order equation, 
\be       \sum_{k=0}^n       Z_k\,  z^{n-k}  =   Z_0   \, \prod_{\ell=1}^n   (z-  w_{\ell})  =     0\;.      \label{stellar1}  
\ee
As the equation (\ref{stellar1}) is invariant under the rescaling  (\ref{rescal}), the collection of the $n$ roots for $z$,
$\{ w_{\ell}\}$, can  each  be  regarded as equivalent to a point of  $\mathbf{CP}^{n}$.    Note that no generality is lost by considering the neighborhood $Z_0 \ne 0$ above:  for instance, one may introduce the variable  $z^{\prime} \equiv 1/z$ and rewrite equation (\ref{stellar1}) in terms of $z^{\prime}$.  One sees then that the equation  with  $Z_0 =0$ contains (at least one) root $z^{\prime} =0$, that is   $z=\infty$. 

\begin{figure}
\begin{center}
\includegraphics[width=4in]{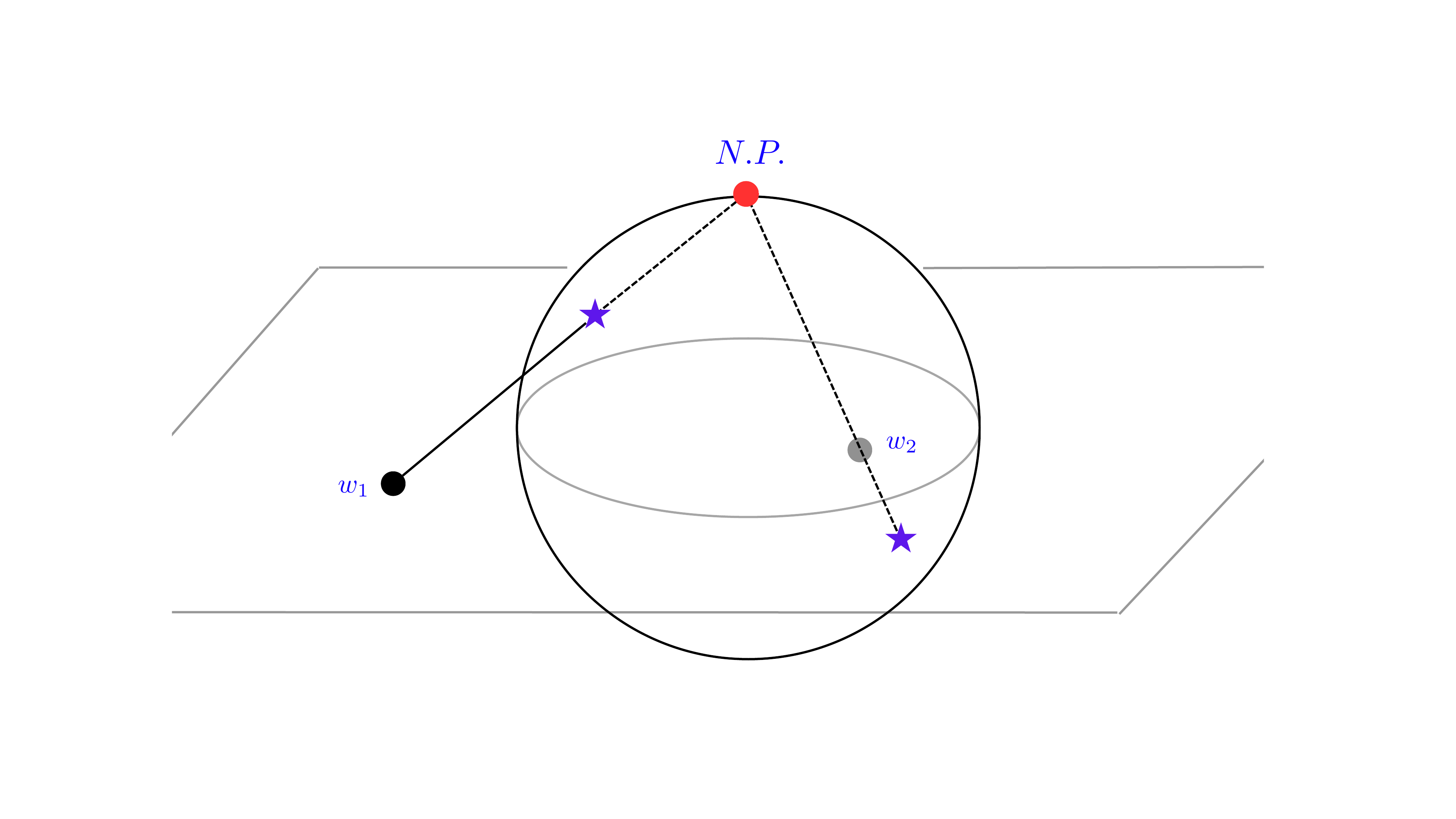}
\caption{\footnotesize   Two points (stars) on the sphere, and their  stereographic projections from the north pole (N.P.)  onto the plane containing the equator.   }
\label{Stereo}
\end{center}
\end{figure}

\begin{figure}
\begin{center}
\includegraphics[width=4in]{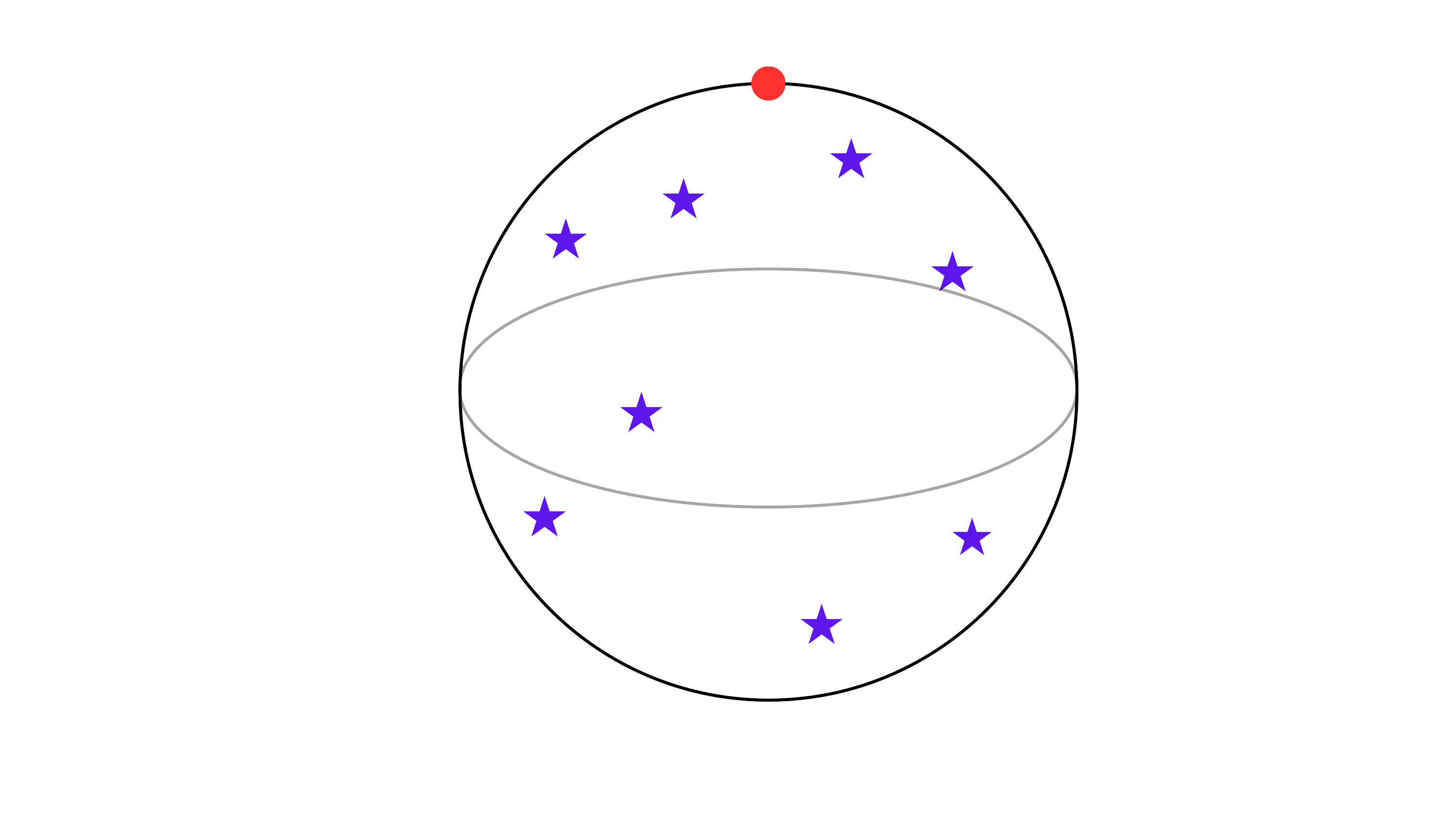}
\caption{\footnotesize  A general point in  $\mathbf{CP}^{n}$ is equivalent to an unordered collection of  $n$  points (stars) on a sphere,   in the stellar representation, (\ref{stellar1}). Each star is associated with the complex number $w_{\ell}$ by a stereographic projection, as in Fig.~\ref{Stereo}. }
\label{Stars}
\end{center}
\end{figure}

The explicit relation between $\{ w_{\ell}\} $ and the $\mathbf{CP}^{n}$ point  can be found by  working backward:    given any  collection of  $n$ roots $\{ w_{\ell}\} $
in the complex plane,  one can find the coefficients  $\{ Z_k \} $ - the coordinates of   $\mathbf{CP}^{n}$ -  by simply   expanding the second expression of (\ref{stellar1}).  
The connection between this construction and the picture (\ref{young}) above  comes from the fact that any complex number $\{ w_{\ell}\} $  can be regarded as the stereographic projection of a point on a sphere $S^2$, from the north (or south) pole, onto the plane containing the equator.  See Fig.~\ref{Stereo}.  Naturally, a point on $S^2$ can be identified with the unit vector ${\mathbf n}$, (\ref{stella0}), the orientation of each component  spin $\frac{1}{2}$.  The explicit relation is
\be   w_{\mathbf n}   =     \frac{e^{i\phi} }{\tan (\theta/2)}\;.   \label{explicit}
\ee
Collecting things, we see that  a point in   $\mathbf{CP}^{n}$ is represented by an unordered collection of  $n$   
points on a sphere (stars), see Fig.~\ref{Stars}.  This is known as the stellar representation of   $\mathbf{CP}^{n}$.   

From the group-theoretic point of view, (\ref{young}), 
a generic quantum state of spin $j$,   is equivalent to the collection (the direct product) of  $n=2j$ spin $\frac{1}{2}s, $
\be      {\cal S} \,\,  |{\bf n}_1 \ckt |{\bf n}_2 \ckt \ldots  |{\bf n}_n \ckt  \label{This}
\ee
where ${\cal S} $  stands for a symmetrization, 
  orientated towards various directions  ${\mathbf n}_k$,   $k=1,2,\ldots, n$,   see  Eq.(\ref{WFspin120}).   (\ref{This}) 
represents a general state of spin $j= 2n$ and naturally corresponds to the random sets of stars in Fig.~\ref{Stars}.   

\begin{figure}
\begin{center}
\includegraphics[width=4in]{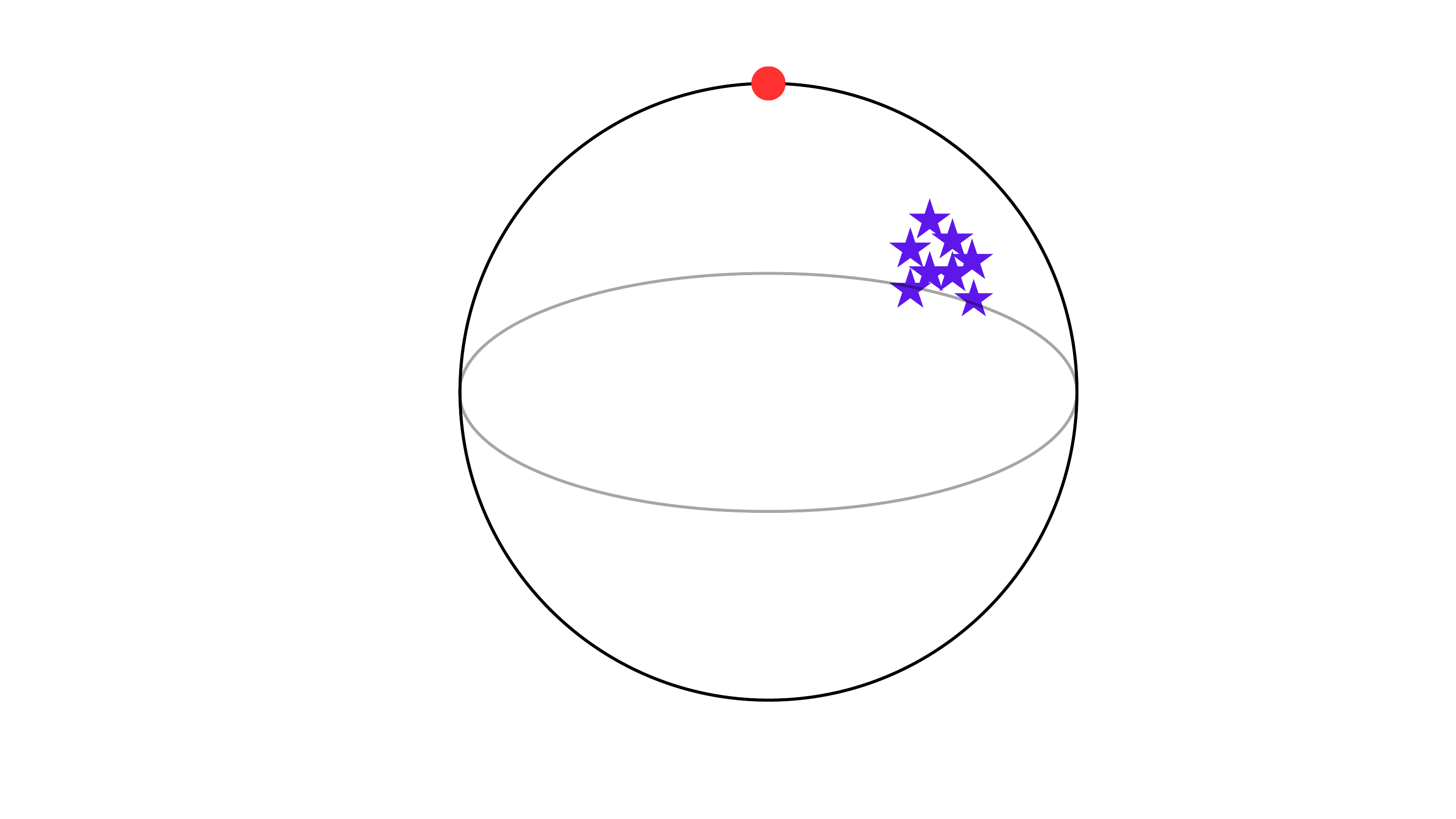}
\caption{\footnotesize  The coherent spin state (\ref{Bloch2}), (\ref{see00}), (\ref{exercise00})   in the stellar representation.     }
\label{coincident}
\end{center}
\end{figure}

  The  particular state in which  the spin  $j$  is oriented towards a definite direction ${\mathbf n}$,    (\ref{Bloch2}),   corresponds to the  case in which the component $\frac{1}{2}$ spins are  all oriented towards the same direction  
${\mathbf n}_1={\mathbf n}_2= \ldots = {\mathbf n}_n=  {\mathbf n}$,
\be          |j, {\mathbf  n}\ckt  =        |{\bf n} \ckt |{\bf n} \ckt \ldots  |{\bf n} \ckt     \label{thesame} 
\ee
(which is automatically symmetric).  See Fig.~\ref{coincident}.    By using  (\ref{WFspin120}) for each  $ |{\bf n} \ckt $, and collecting terms with the same fixed number $k$   of the spin up factors, we find $|j, {\mathbf  n}\ckt$ as an 
expansion in terms of the  eigenstates of $J_z$ with $J_z=m$,   
\be     
|j, {\mathbf  n} \ckt  = \sum_{k=0}^{2j}       c_k \,   |j, m \ckt\;,  \qquad  m = -j+k \;,    \label{see00} 
\ee
\be   
c_k =      {\binom{2j}{k}}^{1/2}   \,     e^{i (j-k) \phi } \left(\cos\tfrac{\theta}{2}\right)^{k}   \left(\sin\tfrac{\theta}{2} \right)^{2j -k} \;, \qquad   \sum_{k=0}^{2j}   |c_k|^2 =1\;, \label{exercise00}
\ee   
where  $ \binom{n}{k} = n!/k!(n-k)!$ are the binomial coefficients.  The states (\ref{see00}), (\ref{exercise00}),  coincide, apart from an overall phase factor, with  the Bloch (or spin coherent) states considered in many-body physics context  \cite{Radcliffe,Puri, Arecchi,Aravind,Lieb,Wodk,Livine,LohKim,Byrnes}, and  denoted as 
$|j, \Omega\ckt$,   $|j, \{\theta, \phi\} \ckt$, etc.

In order to understand this formula from the point of view of the stellar representation of  $\mathbf{CP}^{n}$,   we need to translate the homogeneous coordinates of  $\mathbf{CP}^{n}$ to the coefficients $\{c_k\}$  in  (\ref{see00})   as
\be         Z_k   =   c_k^2\;,     \qquad  k=0,1,\ldots, n\;,   \label{replace1}
\ee 
and at the same time replace the projection of the stars on $S^2$ as
\be      w_{\ell}  \to     w_{\ell}^2\;, \qquad  \forall \ell\;.   \label{replace2}
\ee
Eq.(\ref{replace1}) and (\ref{replace2})  reflect the double covering of SO(3) group by SU(2).

In the stellar representation of  $\mathbf{CP}^{n}$,  (\ref{rescal}), (\ref{stellar1}),  the state (\ref{thesame}) corresponds to the situation where all roots coincide, 
\be       w_\ell=  w_{\mathbf n}   \;, \qquad \forall \ell\,.   
\ee
All stars are at a coincident point  ${\mathbf n} $    on the sphere, see Fig.~\ref{coincident}.
Taking into account the doubling (\ref{replace1}) and (\ref{replace2})  and 
expanding  ($n=2j$) 
\be
\sum_{k=0}^n       (c_k)^2  \,  z^{n-k}    =   \prod_{\ell=1}^n   (z -  w_{\ell}^2)  = (z- w_{\mathbf{n}}^2)^n  \;,
\ee
($w_{\mathbf n}   =  \frac{e^{i\phi} }{\tan (\theta/2)}$),   
one finds precisely  (\ref{see00}), (\ref{exercise00}), apart from unobservable phases and an overall factor.

Thus the stellar representation of $\mathbf{CP}^{n}$ points  - the generic states of spin $j$ -  provides us  with an appealing  intuitive picture of quantum fluctuations.  The  $\mathbf{CP}^{n}$ points  with randomly distributed stars, Fig.~\ref{Stars},  represent  spin states with large quantum fluctuations.
The special states,   (\ref{Bloch2}), (\ref{thesame}) - (\ref{exercise00}) with coincident stars, Fig.~\ref{coincident},  instead, are those with the minimum  fluctuations,  and  correspond to  the Bloch or spin coherent state   \cite{Radcliffe, Puri, Arecchi, Aravind, Lieb}.
In the next four sections we shall  illustrate  how these states effectively behave as the classical angular momenta,   $j \, {\mathbf n}$,    in the $j\to \infty$  limit.

\section{Stern-Gerlach experiments   \label{SG}  } 

In this section  we compare the Stern-Gerlach experiment  \cite{SG}  for small and large spins, with special attention to the states,  
(\ref{Bloch2}), (\ref{see00}), (\ref{exercise00}). 

\subsection{Spin \texorpdfstring{$1/2$}{1/2} atom   \label{SG1}}

For definiteness let us take an incident atom with spin $j = \tfrac{1}{2}$   directed along a definite, but generic, direction ${\bf n}=  (\sin \theta  \cos \phi,  \sin \theta  \sin \phi,   \cos \theta)$, i.e.
\be  
\Psi =   \psi({\bf r}, t)  \, |{\bf n} \ckt \;, \label{WF}
\ee
\be      
|{\bf n} \ckt    =    c_1 |\!\uparrow\ckt +  c_2 |\!\downarrow\ckt \;,   \qquad  c_1=   e^{-i \phi/2}  \cos \tfrac{\theta}{2}\;, \qquad c_2=   e^{i \phi/2}  \sin \tfrac{\theta}{2}\;.   \label{WFspin12}  
\ee
Let us suppose that the atom enters from $x=-\infty$ and moves towards $x=+\infty$.    
When the atom passes through a region with an inhomogeneous  magnetic field with  
\be
  \frac{\de B_z}{\de z} \ne 0\;,   \label{Bz}
\ee
the wave packet splits into two subpackets,  with  respective weights, $|c_1|^2$ and  $|c_2|^2$:  they manifest as the relative intensities of the two separate image bands on the screen at the end.  

Even though the Stern-Gerlach process is discussed in every textbook on quantum mechanics, there is a subtlety, relatively little known.  The (apparent) puzzle is why  the net  effect is the deflection of the atom towards the $\pm {\hat z}$ direction only,    in spite of the fact that  Maxwell's equations $\nabla \cdot {\bf B}=0,\,\, \nabla \times {\bf B}=0$ dictate  that the inhomogenuity $(\ref{Bz})$ implies an inhomogenuity ${\de B_y}/{\de y}\,$ of the same magnitude (assuming $B_x=0$).
The explanation,  with the discussion on the characteristics of the appropriate magnetic fields,  has been given in \cite{Platt,Alstrom}.  We briefly review this story in Appendix~\ref{SGB}.
 
The original Stern-Gerlach experiment (1922) \cite{SG}   has had a fundamental impact to the development of quantum theory,  in proving the quantization of the angular momentum, the existence of half-integer spin, and more generally, demonstrating the quantum-mechanical nature of the silver atom as a whole \cite{KKHTE}.     

The wave function of the form (\ref{WFspin12}) corresponds to a pure state,  i.e., $100$\%   polarized beam  where all incident atoms are in the  same  spin state. 
If the beam is partially polarized or unpolarized,  the state is described by a generic  density matrix $\rho$, with  
\be   \Tr \, \rho =  1\;, \qquad   \rho_{ii}  \ge 0\;, \qquad i=1,2\;.
\ee

The  Stern-Gerlach  experiment measures the relative frequency that the atom arriving at the screen happens to have spin $s_z=  \tfrac{1}{2}$ or 
 spin  $s_z=  -\tfrac{1}{2}$.  Let 
 \be   \Pi_{\uparrow} =  |\uparrow\ckt\brc \uparrow|  = \left(\begin{array}{cc}1 & 0 \\0 & 0\end{array}\right) \;, \qquad    \Pi_{\downarrow} =  |\downarrow\ckt\brc \downarrow|    = \left(\begin{array}{cc}0 & 0 \\0 & 1\end{array}\right)   \ee
 be the projection operators on    the spin up (down) states;    
 the relative intensities of the upper and lower blots on the screen are,   according to QM,  
\be    {\bar \Pi_{\uparrow} } =  \Tr   \Pi_{\uparrow}  \rho =  \rho_{11}\;;  \quad    {\bar \Pi_{\downarrow} } =  \Tr   \Pi_{\downarrow}  \rho =   \rho_{22}\;, \qquad 
 \rho_{11}\ +  \rho_{22}=1\;.    \label{density}     \ee
The prediction about the relative intensities of the two narrow atomic image bands  from the wave function  (\ref{WFspin12}) and from the density matrix, 
(\ref{density}),  is in general indistinguishable, as is well known.

\subsection{Large spin    \label{SG2}}

 Consider now  the state of a spin $j$ directed towards a direction  $ {\mathbf  n}$,   $ |j,  {\mathbf  n}\ckt $,  given  in 
 (\ref{Bloch2}),(\ref{see00}), and (\ref{exercise00}),   Fig.~\ref{coincident}.  
Choosing the magnetic field (and its gradients) in the ${z}$-direction, we need to express  $ |j, {\mathbf  n}\ckt $ as a superposition of the eigenstates of ${\hat  {J_z}}$  ($j= n/2$). This is precisely the expression  given in Eq.(\ref{see00}) and Eq.(\ref{exercise00}).

\begin{figure}
\begin{center}
\includegraphics[width=4.5 in]{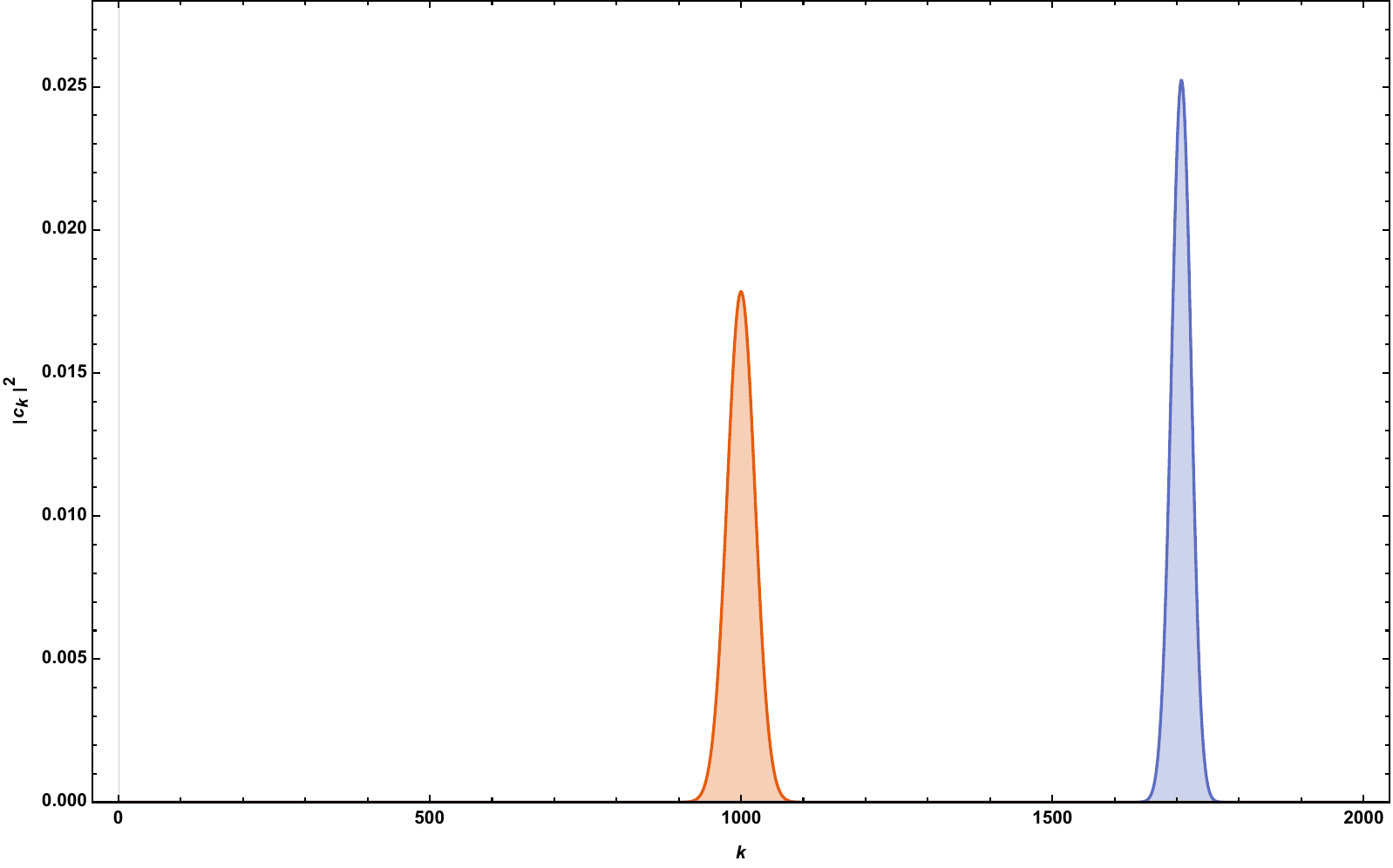}
\caption{\footnotesize   A distribution in possible values of $J_z=-j +k$,  for a spin $j  =1 \cdot 10^3 $ particle in the state (\ref{see00}),  (\ref{exercise00}), with $\theta = \pi/2$ (center, orange peak) or  with  $\theta = \pi/4$  (right, blue peak). 
 The graph represents the distribution $|c_k|^2$. 
The particle starts looking like classical.} 
\label{Spreadno}
\end{center}
\end{figure}

\begin{figure}
\begin{center}
\includegraphics[width=4.5 in]{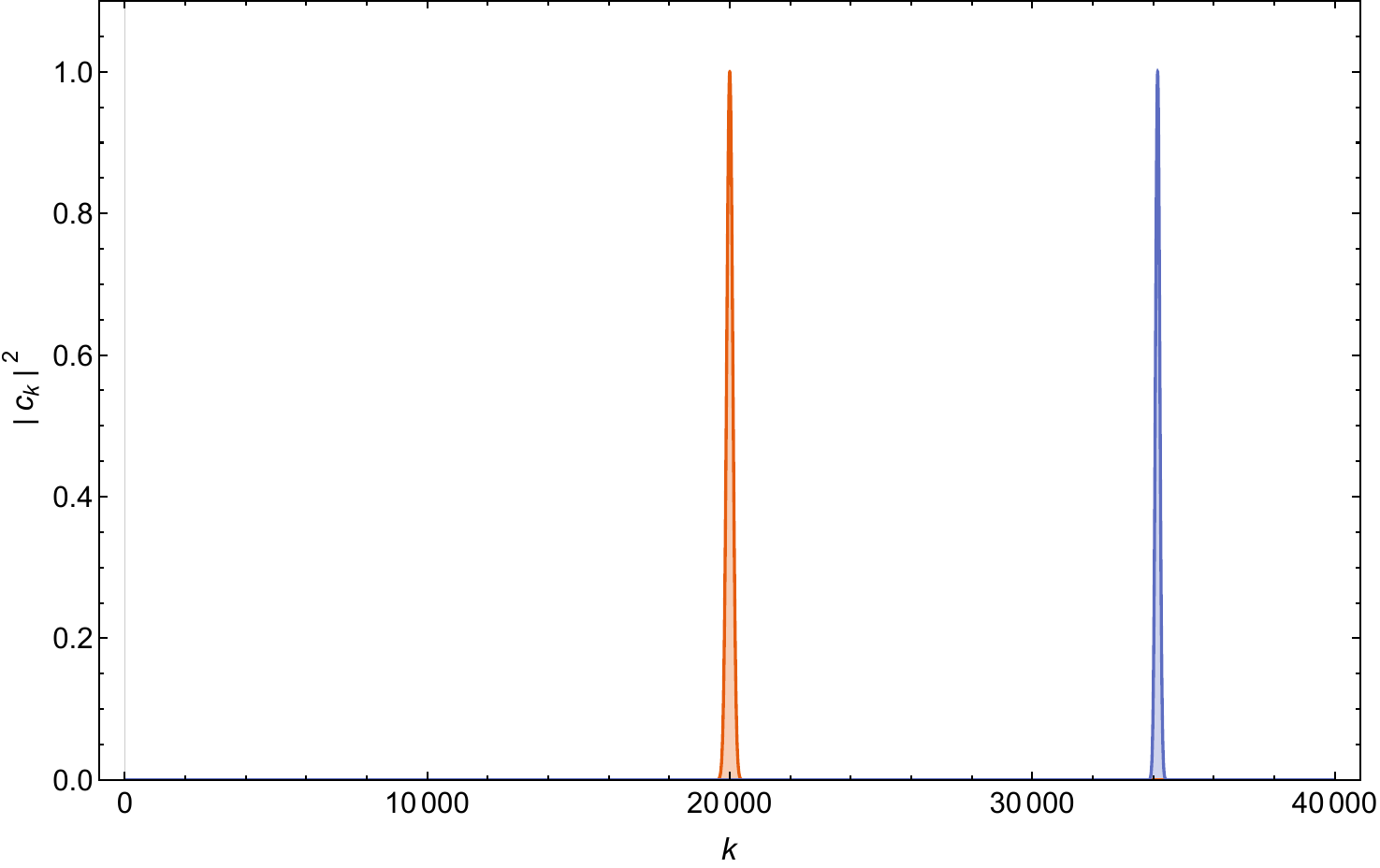}
\caption{\footnotesize    The distribution in $J_z$  ($\equiv -j +k$) as in Fig.~\ref{Spreadno} but for a spin $j  =2 \cdot 10^5$,  for $\theta = \pi/2$ (center, orange peak) or  $\theta = \pi/4$  (right, blue peak).  This figure is drawn by using the approximation (\ref{distr}) and  (\ref{distrLAM}), rather than the exact formula,  (\ref{see00}), (\ref{exercise00}). }
\label{Spreadno2}
\end{center}
\end{figure}

Using Stirling's formula in (\ref{exercise00}), one finds, for  $n$ and $k=j+m$  both large  with $x = k/n$ fixed, the following  distribution in different values of $ m=-j+k$, 
 \be 
|c_k|^2   \simeq    e^{n  f(x)}\;, \qquad   x= k/n=(j+m)/2j \;,    \label{distr}    
\ee
where
\be  
 f(x)   =  - x \log x - (1-x) \log (1-x)    + 2 x  \log  \cos \tfrac{\theta}{2}  + 2 (1-x) \log   \sin \tfrac{\theta}{2}  \;.                \label{distrLAM}  
\ee
The saddle-point approximation  valid at $n \to \infty$, yields
\be   
f(x)    \simeq     - \frac{(x-x_0)^2 }{x_0(1-x_0)}   \;,  \qquad      x_0  =     \cos^2  \tfrac{\theta}{2}\;,  \label{spike1} 
\ee
and therefore 
\be    
\sum_k   |c_k|^2  (\cdots )     \longrightarrow      \int_0^1    dx   \,    \delta(x-x_0)  \, (\cdots )   \label{spike2} 
\ee
in the $n \to \infty$  ($x= k/n$ fixed)  limit. The narrow peak position $x=x_0$  corresponds to (see  Eq.(\ref{see00})) 
\be  
J_z =m  = n  (x-\tfrac{1}{2}) =   j  \, (2  \cos^2  \tfrac{\theta}{2}-1)  =   j \, \cos \theta\;.   \label{selection}
\ee
This means that a large spin ($j  \gg \hbar$) quantum particle with spin directed towards ${\mathbf n}$, in a Stern-Gerlach setting with an inhomogeneous magnetic field (\ref{Bz}), moves along a single trajectory of a classical particle with  $J_z=   j  \, \cos \theta$,  instead of spreading over a wide range of split sub-packet trajectories covering  $-j  \le m   \le j$.   See  Fig.~\ref{Spreadno} (spin $j=10^3$)  and   Fig.~\ref{Spreadno2} ($j=2 \cdot 10^5$)  and compare them with  Fig.~\ref{Spreadsi}  for spin $j=13/2$.  

This (perhaps) somewhat surprising result appears to indicate  that  quantum mechanics (QM)  takes care of itself,  so to speak,  in ensuring that a large spin particle   ($j/\hbar \to \infty$)  behaves classically, at least for these particular states $|j, {\mathbf n}\ckt $.     No extra conditions are necessary.  This is consistent with the  known general behavior  of the wave function in the semi-classical limit  ($\hbar \to 0$)  \footnote{Of course this does not mean that the classical limit necessarily
requires or implies $j/\hbar \to \infty$.}.
Note that  if the value  $j   \gg \hbar $  is understood as due to the large number of spin $\frac{1}{2}$ particles composing it (see the comment after  (\ref{exercise00})),  the spike  (\ref{spike1}),   (\ref{spike2}) can be  understood as due to the accumulation of a huge number of microstates giving $J_z  \simeq  m$.   

Another, equivalent way to see the shrinkage of the distribution in $J_z=m$, (\ref{spike2}), is to study its dispersion. Given the relative weights $|c_k|^2$, (\ref{exercise00}),  where  $m =  - j + k$ and $j  =  \tfrac{n}{2}$, with $k=0,1,\ldots, n$, it is a simple exercise to find that 
\be     
\brc  J_z \ckt =  j   \, \cos \theta\;, \qquad  \brc  J_z^2 \ckt  =  j^2 \cos^2 \theta +  \frac{j}{2}  \sin^2 \theta\;,   \label{fluc1} 
\ee
so that the dispersion (the fluctuation width) is  given by \footnote{These results are well  known  \cite{Puri}.}
\be    
(\Delta J_z)^2 =    \brc     (J_z  -    \brc  J_z \ckt )^2   \ckt =    \brc  J_z^2  \ckt -   \brc  J_z \ckt^2   =    \frac{j}{2}  \sin^2 \theta\;.    \label{fluc2} 
\ee
It follows that 
\be  
\lim_{j /\hbar  \to \infty}  \frac  {|\Delta J_z |}{j}  =0\;,    \label{fluc3} 
\ee
i.e. it is an infinitely narrow distribution. Note that the SG experiment is indeed a measurement of $J_z$ in the state (\ref{see00}). The fact that the result is always  $J_z\sim \brc J_z \ckt =  j  \, \cos \theta$  with no dispersion, means that it is a classical angular momentum  $\sim j\,  {\mathbf n}$.

%
%
%

\subsubsection{The state preparation and successive SG set-ups\label{successiveSG}   }      
For a small $j$, as it is well known, the SG set-up can be used to prepare any desired state $|j, J_z=m \ckt$, by an intentionally biased SG apparatus, e.g. in which all the split sub-wavepackets are blocked, except  for the one corresponding to  $|j, m\ckt$. Equivalently,  one may use the so-called null-measurements (see Ref.~\cite{KK3} for a recent review). In any case, for a small finite $j$, where there are $2j+1$ well-separated sub-packets (see Fig.~\ref{Spreadsi}), there is no difficulty in extracting any desired quantum state for subsequent experiments (state preparation).

 In the large-spin limit ($j \to \infty$) of the Bloch states  $ |j, {\mathbf  n}\ckt $  we are considering, there is no possibility of extracting or selecting a desired sub-wavepacket,  since there is essentially only one narrow wavepacket present (see (\ref{selection}),  Fig.~\ref{Spreadno2}). Trying to extract a state with 
 $J_z$ not close to  $ j  \cos \theta $ would simply give a zero result. 
 We have in mind a macroscopic scenario where, in comparison to the finite region (in principle) occupied by the waves $-j \leq m \leq j$, the fluctuations of the values of $m$ around (\ref{selection}) are of the order of $\sqrt{j} \ll  j  $, i.e. of zero width.

We might consider introducing a second set of Stern-Gerlach (SG) magnetic fields directed in a different direction, denoted by ${\hat z}^{\prime}$, as a new quantization axis. This setup allows us to study the state emerging from the first set of SG magnetic fields (but without an imaging screen),
\be     
|j, {\mathbf  n}\ckt  \sim  c_k \,   |j, J_z= m \ckt\;,  \qquad   m    \sim  j \, \cos \theta\;.   \label{seeNow} 
\ee
Since the sum of states around $m  \sim  j \, \cos \theta$ is essentially equal to the complete sum, this state can be rewritten as
\bea     
 |j, {\mathbf  n}\ckt    & = &    \sum_{m, m^{\prime}}    c_k \,    |j,  m^{\prime}  \ckt   \brc   j,  m^{\prime}  |j,  m \ckt \nonumber \\
 &=&     \sum_{m, m^{\prime}}   |j,  m^{\prime}  \ckt   \brc   j,  m^{\prime}    |j,  m \ckt     \brc j,  m | j, {\mathbf  n}\ckt    \nonumber \\ 
 &=&    \sum_{m^{\prime}}   |j,  m^{\prime}  \ckt   \brc   j,  m^{\prime}  | j, {\mathbf  n}\ckt   =    \sum_{m^{\prime}}    c_{k^{\prime}} |j,  m^{\prime}  \ckt   \;,
\label{SGbis}  \eea
where in the first line a completeness relation $\sum_{m^{\prime}}    |j,  m^{\prime}  \ckt   \brc   j,  m^{\prime} |={\mathbf 1}$ has been inserted, and in the second   we recalled   $c_k= \brc   j,  m  | j, {\mathbf  n}\ckt $ (see Eq.(\ref{see00})).  
(\ref{SGbis})  represents an expansion of the original state $|j, {\mathbf n}\rangle$ as a linear combination of the eigenstates of $J_{z^{\prime}}=m^{\prime}$. In the limit $j \to \infty$, the results from Section~\ref{SG2} tell us that the sum is dominated by the small region around $m^{\prime} \sim j \cos ({\mathbf n} \cdot {\hat z}^{\prime})$, with negligible fluctuations. The trajectory remains unique, regardless of the choice of quantization axis (i.e. the direction of the magnetic field) ${\hat z}^{\prime}$.

\subsubsection{States almost directed towards \texorpdfstring{$\vec{\mathbf{n}}$}{n}}  \label{almostOr}
Obviously, in the  large spin limit,  other states close to  $ |j,  {\mathbf  n}\ckt $,  
 \be   
 \left({\hat {\mathbf  j}}  \cdot {\mathbf  n} \right) \,  |j,  {\mathbf  n}, q \ckt   = ( j-q)\,  |j,  {\mathbf  n}, q\ckt \; ,     \qquad  q=1,2, \ldots  \,\,\, {\rm fixed}  \label{almost}  
 \ee 
(i.e. the spin roughly oriented towards  ${\mathbf n}$) will behave similarly. For instance, the  $q=1$ state is (cfr. Eq.(\ref{see00}))  
\bea     
|j, {\mathbf  n}, 1\ckt  &=& \sum_{\ell=0}^{N-1}   \frac{1}{\sqrt{N}}  \,  |{\bf n} \ckt_1 \otimes \cdots   |{\bf n} \ckt_{\ell}   \otimes   |-  {\bf n} \ckt \otimes 
|{\bf n} \ckt_{\ell+2} \otimes \cdots   |{\bf n} \ckt_{N}   \;, \nonumber  \\
&=&   \sum_{\ell=0}^{N-1}     \frac{1}{\sqrt{N}}      |j_1, {\mathbf n} \ckt     \otimes   |-  {\bf n} \ckt \otimes      |j_2,  {\mathbf n} \ckt \;,    \label{seeBis} 
\eea
where  $  |j_1, {\mathbf n} \ckt$ and  $  |j_2, {\mathbf n} \ckt$  ($j_1 = \tfrac{\ell}{2} \;,  j_2= \tfrac{N- \ell-1}{2}$),  are two spins oriented towards  ${\mathbf n}$,  (\ref{Bloch2}),  as already studied in  Sec.~\ref{SG2}.  In the $j \to \infty$ limit, the sum is dominated by terms with $\ell \sim \mathcal{O}(N) \to \infty$.  The SG  projection of  this state onto the $j_z$ eigenstates is therefore approximated by  
 \be       
 j_z \sim   j_{1\, z} +    j_{2\, z} =   j_1 \cos \theta + j_2 \cos \theta \simeq  j\, \cos \theta\;, 
 \ee
where the results of Sec.~\ref{SG2} have been used.   A similar result holds for  $|j,  {\mathbf  n}, q \ckt $  for any finite $q$, showing that all these states  behave as a classical angular momentum, $j\, {\mathbf n}$.

\section{Orbital angular momentum \label{Orbital} } 
All our formal arguments (\ref{commutation})--(\ref{corresp3}) apply equally well, both to spin (the intrinsic angular momentum of a particle or the total angular momentum of the center of mass of a body in its rest frame) and to orbital angular momentum. However, our primary tool of analysis is the Stern-Gerlach (SG) set-up (Sec.~\ref{SG2}), which is more suitable for spin than for an angular momentum associated with the orbital motion of a particle. One might wonder how the large angular momentum limit works for the latter. Here is a brief acccount.

\begin{figure}[!htb]
    \centering
    \begin{minipage}{0.4\textwidth}
        \centering
        \includegraphics[width=1.0 \linewidth, height=0.2\textheight]{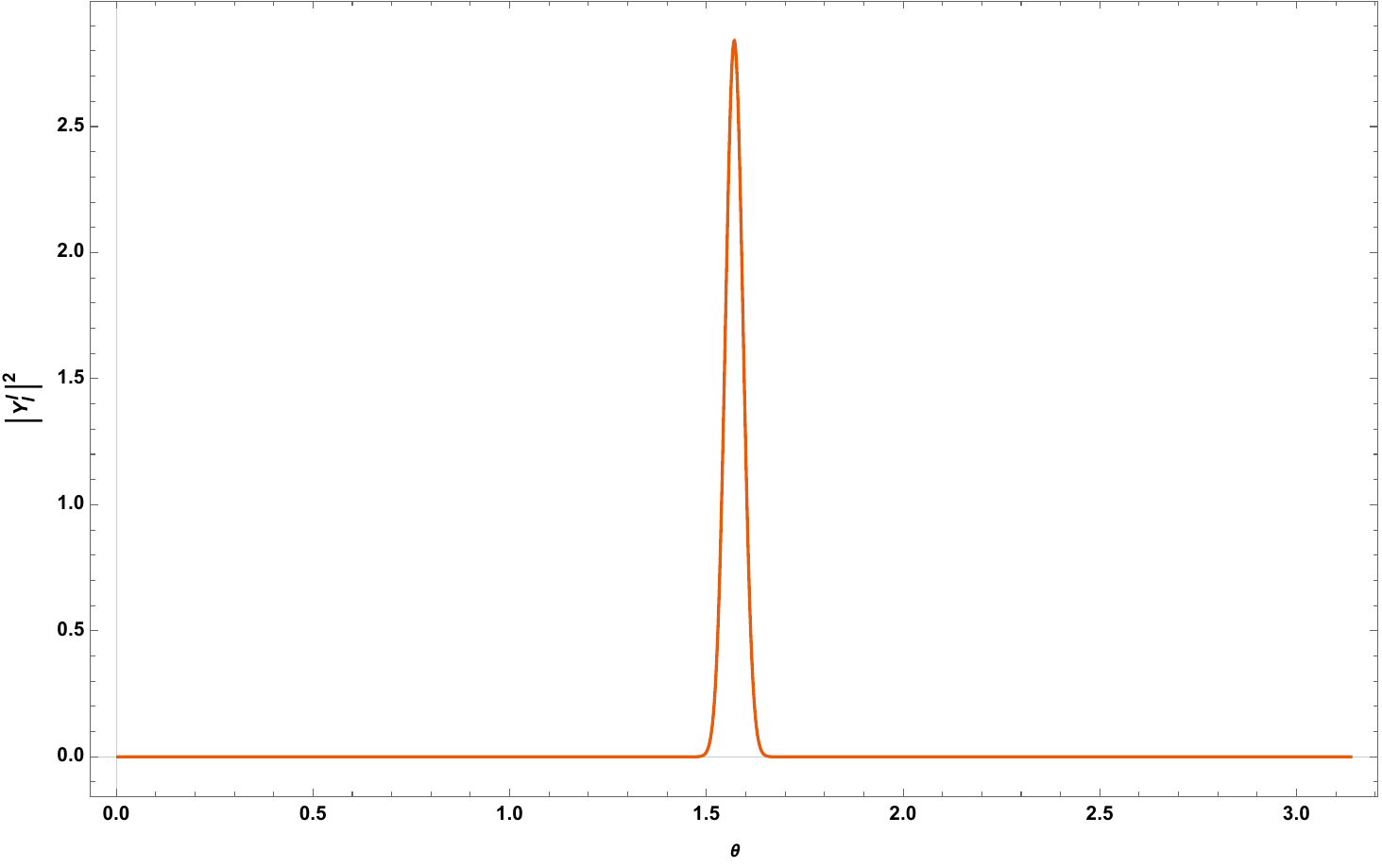} 
        \caption{\footnotesize The distribution in $\theta$,  for \\ the state   $| \ell, \ell\ckt $ with $\ell=10^3$.   }
        \label{ell1000}
    \end{minipage}
    \hspace{0.05\textwidth}
    \begin{minipage}{0.4\textwidth}
        \centering
        \includegraphics[width=1.0\linewidth, height=0.2\textheight]{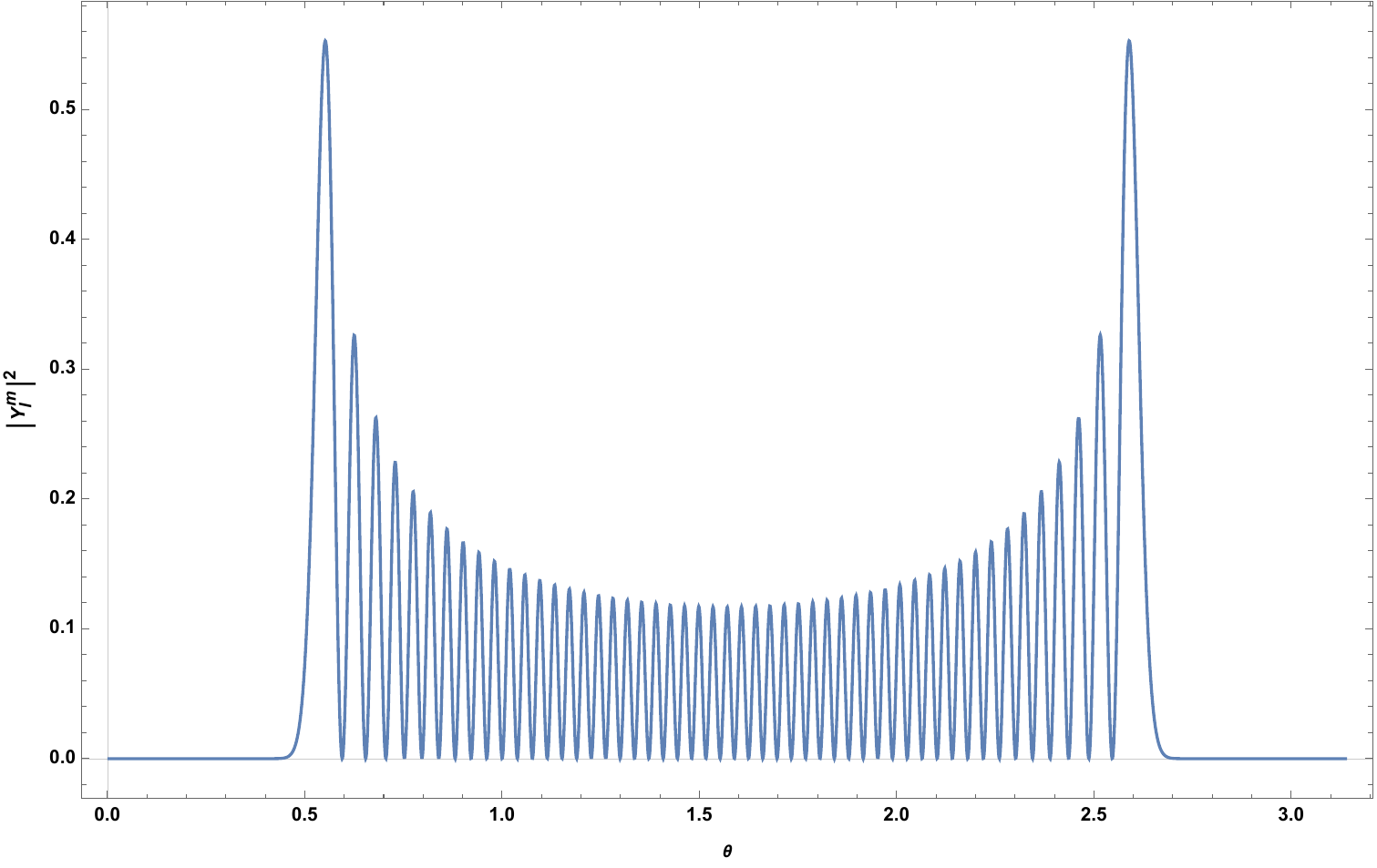}
        \caption{\footnotesize The distribution in  $\theta$, for \\ the state $| \ell, m \ckt $ with $\ell=100$, $m=50$.  }
        \label{ell100}
    \end{minipage}
\end{figure}

 The (angular part of the) wave function of a particle in the state $|\ell, m\rangle$ corresponds to the spherical harmonics,
\be   
\brc \theta, \phi |  \ell, m\ckt =    Y_{\ell,m}(\theta, \phi) =  (-)^{(m+|m|)} \sqrt{\tfrac{(2\ell+1)(\ell-|m|)!}{4\pi (\ell+|m|)!}} P_{\ell}^{|m|} (x) \, e^{i m \phi}\;,\qquad x \equiv \cos \theta\;,
\ee
where  $P_{\ell}^m(x)$  are the associated Legendre polynomials. Let us consider the state of minimum fluctuation,  $| \ell, \ell\ckt $  (see Eq.(\ref{minimum})).  It is easy to see that the angular distribution is strongly peaked at $\theta = \tfrac{\pi}{2}$ at large $\ell$,
\be     
|Y_{\ell,\ell}(\theta, \phi)|^2 =   \frac{2 \ell+1}{4\pi} \frac{(2\ell)!}{2^{2\ell} (l!)^2}   (1-x^2)^{\ell}  \stackrel{\ell \to \infty}{\longrightarrow}   \frac{1}{2\pi}  \delta(x) \;. \label{delta} 
\ee
For instance, see Fig.~\ref{ell1000}  for $\ell=m= 10^3$.  In states $| \ell, m\ckt $ with stronger fluctuations, with $m$ not close to $\pm \ell$,   the distribution instead covers a large portion of the $3$D solid angle,  for any $\ell$  (see Fig.~\ref{ell100} for $\ell=100,  m=50$). 

Now a classical angular momentum has its all three components well defined. By choosing an appropriate coordinate system,  it can be written as  ${\bf L}^{(\rm class)} = (0,0, L)$. It describes a particle moving in the $(x,y)$ plane, with  $z=p_z=0$.   The large $\ell$ limit of the wave function $ |Y_{\ell,\ell}(\theta, \phi)|^2$,  (\ref{delta}), describes precisely such a motion.  

\section{Addition of angular momenta  \label{Addition}}  
Let us now  consider two spins,   ${\mathbf j}_1$ and  ${\mathbf  j}_2$. The composition-decomposition rule in QM is well known, i.e.
\be  
 {\underline{j_1}} \otimes   {\underline{j_2}} =       {\underline{j_1 + j_2}} \oplus    {\underline{j_1 + j_2-1}}  \oplus  \ldots     {\underline{{|j_1 - j_2|}}} \;,  \label{composition} 
\ee
where $\underline{j}$  indicate the multiplet  (the irreducible representation) of  SU(2) \footnote{We recall that SU(2) is the simply-connected  double cover of the orthogonal group SO(3).}, with multiplicity $2j+1$ and with $\max \{j_z\}=j$. We set $\hbar=1$ here.

We wish to find out how the addition rule looks like in the limit  $j_1, j_2 \to \infty$.   
We are particularly interested in  
the composition rule for angular momentum states  with minimum fluctuations,   of the form,  $|j, {\mathbf n}\ckt $.  
Therefore, let us consider two particles (spins) in the states,
\be     
{\hat {\mathbf j}}_1 \cdot {\mathbf n}_1 \,   |j_1, {\mathbf n}_1 \ckt  =     j_1 \,  |j_1, {\mathbf n}_1 \ckt \;,   \qquad    {\hat {\mathbf j}}_2 \cdot {\mathbf n}_2 \,  |j_2,  {\mathbf n}_2 \ckt  =     j_2 \,  |j_2, {\mathbf n}_2 \ckt \;,   \label{prepare}
\ee
namely, they are spins oriented towards the directions,   $ {\mathbf n}_1$ and  $ {\mathbf n}_2$, respectively.   Our aim is to find out the properties of the direct product state,
\be        
|j_1,  {\mathbf n}_1 \ckt \otimes    |j_2,  {\mathbf n}_2 \ckt   \;,\label{product}  
\ee
in the large $j_1, j_2$ limit. 

Since the choice of the axes is arbitrary, we may take 
\be  
{\mathbf n}_2 = (0,0,1) \;.  
\label{reference} 
\ee
Then the product state is just 
\be     
|j_1,  {\tilde  {\mathbf n}}_1 \ckt \otimes       |j_2,m_2=j_2  \ckt \;,
\ee
where  $ |j_2, m_2=j_2  \ckt $ is the highest  $\hat{j}_{2\, z}$ state.  We added the tilde sign on  $ {\mathbf n}_1$ to indicate that it is the vector  ${\mathbf n}_1$ in the reference system,   (\ref{reference}). Now  $ |j_1,   {\tilde  {\mathbf n}}_1 \ckt $  in the $j_1 \to \infty$ limit has been studied in Sec.~\ref{SG2}.  It is 
\be  
|j_1,   {\tilde  {\mathbf n}}_1 \ckt  =  \sum_{m=-j_1}^{j_1}        c_m \,   |j_1,   j_{1\, z}= m\ckt \;, 
\ee
where the sum is dominated by the values of $m$  around 
\be
m= j_{1\,z} =   j_1 \cos \theta \;, \qquad   \cos \theta =  {\tilde  {\mathbf n}}_{1\, z} =   {\mathbf n}_1\cdot   {\mathbf n}_2\;,   
\ee
with the fluctuations  of $ j_{1\,z}$  which become negligible in the infinite spin limit  (see Fig.~\ref{Spreadno2}). 
For the purpose of the discussion of this section, we introduce the notation
\be |j_1,   {\tilde  {\mathbf n}}_1 \ckt  \simeq      |j_1,  j_{1\, z}= j_1 \cos \theta \ckt\;,
\ee
by using the $\simeq$  sign,  to express this fact.
Since the eigenvalues of $\hat{j_z}$ simply add up in the product state, the expansion of $|j_1, {\mathbf n}_1\rangle \otimes |j_2, {\mathbf n}_2\rangle$ in the expansion in terms of the eigenstates  of $\hat{j}_{{\rm tot}\,z}$ is dominated by terms with $ j_{\rm tot}^{{\tilde n}_2} =    j_1 \cos \theta+ j_2 $, namely, 
\be    
|j_1,  {\mathbf n}_1 \ckt \otimes    |j_2, {\mathbf n}_2 \ckt    \simeq  |j_{\rm tot},   j_{\rm tot}^{{\tilde n}_2} =    j_1 \cos \theta+ j_2  \ckt\;,     \label{proj1}  
\ee
where  $ j_{\rm tot}^{{\tilde n}_2} $ means the component in the direction of ${\tilde n}_2$ of the  (for the moment, unknown) total spin  $j_{\rm tot}$. 
Exchanging $ {\mathbf n}_1 $ and  $ {\mathbf n}_2 $ and repeating the arguments, one finds also  that 
\be    
|j_1, {\mathbf n}_1 \ckt  \otimes    |j_2,  {\mathbf n}_2 \ckt     \simeq          |j_{\rm tot},   j_{\rm tot}^{\tilde n_1}  =   j_1 +  j_2 \cos \theta  \ckt\;.   \label{proj2}  
\ee

But  which quantum angular-momentum state does the product state  (\ref{product})  represent?   Such a state must be compatible with 
the projections (\ref{proj1}) and (\ref{proj2}).    
The answer is that  
\be    
|j_1, {\mathbf n}_1 \ckt  \otimes    |j_2,  {\mathbf n}_2 \ckt     \simeq          |j_{\rm tot},  {\mathbf  n} \ckt\;,    \label{qproduct}  
\ee
with   $j_{\rm tot}$ and the unit vector  ${\mathbf n}$  defined by
\be  
j_{\rm tot}  \,  {\mathbf n} =      j_1 {\mathbf n}_1 +    j_2  {\mathbf n}_2\;.  \label{totalj} 
\ee
Note that this (classical) vector sum determines  both  $j_{\rm tot}$  (the magnitude) and ${\mathbf n}$ (the direction) uniquely.   
The quantum state   $ |j_{\rm tot},  {\mathbf  n} \ckt$  is defined via the correspondence (\ref{Bloch2}), (\ref{proposal})  proposed before.

The proof of the statement (\ref{qproduct})  is as follows. By selecting the directions of the SG magnets directions (the quantization axis) to   ${\mathbf n}_2$ or to  ${\mathbf n}_1$  on the state  $ |j_{\rm tot},  {\mathbf  n} \ckt $,   (\ref{totalj}), one can apply the results of Sec.~\ref{SG1} to get exactly  the results (\ref{proj1}) or (\ref{proj2}), respectively. 
Now, according to the quantum-classical correspondence of Eq.(\ref{proposal}), the state $|j_{\rm tot}, {\mathbf n}\rangle$ (see Eq.(\ref{totalj})) translates into
\be     
{\mathbf j}_{\rm tot}^{(\rm class)}   =     j_1 \, {\mathbf n}_1 +  j_2 \, {\mathbf n}_2\;.  \label{classum} 
\ee
But, in view of (\ref{proposal}),  this is precisely the addition rule of the two classical angular momenta.

\begin{figure}
\begin{center}
\includegraphics[width=4in]{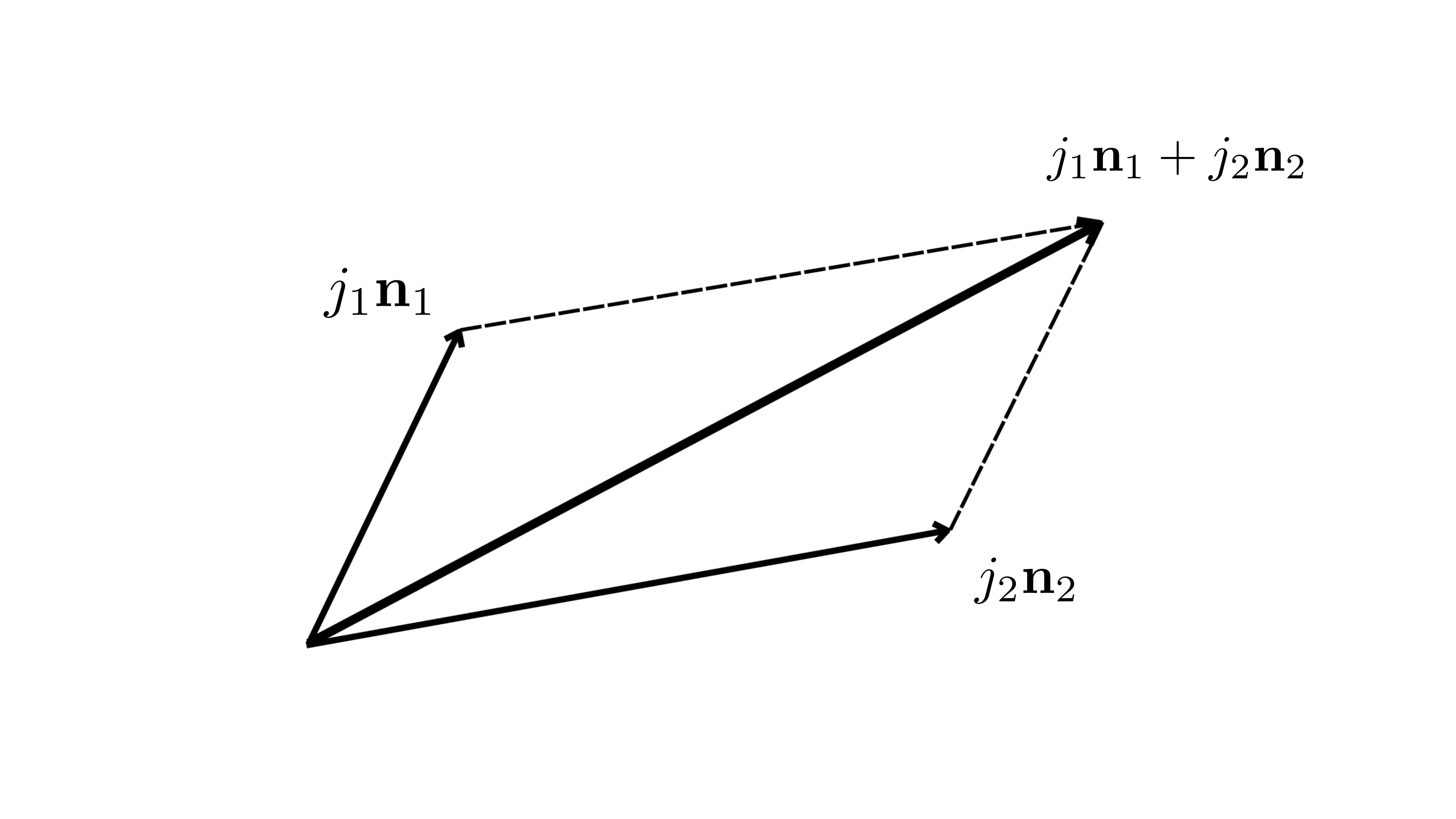}
\caption{\footnotesize   The direct product state, (\ref{product}), behaves as a classical vector sum, in the limit, $j_1, j_2 \to \infty$,
in the sense of the correspondence, (\ref{proposal}). }
\label{VSum}
\end{center}
\end{figure}

To be complete, one must check that the projection on any other  generic direction ${\mathbf n}_3$  works correctly. By using the result of Sec.~\ref{SG2} for the SG projection on a new  ${\mathbf n}_3$  axis,  one finds
\be      
|j_1, {\mathbf n}_1 \ckt \otimes    |j_2, {\mathbf n}_2 \ckt    \simeq         |j_{\rm tot}^{\prime},   j_{\rm tot}^{\prime \, n_3}  =  j_1 {\mathbf n}_1 \cdot {\mathbf n}_3 +   j_2 {\mathbf n}_2 \cdot {\mathbf n}_3   \ckt \;,    \label{again} 
\ee
where we made use of the fact that   ${\hat {\mathbf j}}_1$ and  ${\hat  {\mathbf j}}_2$  commute and the eigenvalues of their  ${\mathbf n}_3$ components  just sum up in the direct product state.  Also, to be conservative, we have left $j_{\rm tot}^{\prime}$ unknown. But
\be    
j_1 {\mathbf n}_1 \cdot {\mathbf n}_3 +   j_2 {\mathbf n}_2 \cdot {\mathbf n}_3    =     ( j_1 {\mathbf n}_1 +   j_2 {\mathbf n}_2)  \cdot {\mathbf n}_3 \;, 
\ee
therefore
\be   
{\mathbf  j}_{\rm tot}^{\prime}=       {\mathbf j}_{\rm tot}  =     j_1 \, {\mathbf n}_1 +  j_2 \, {\mathbf n}_2\;, \label{classumBis} 
\ee
by again using the results of Sec.~\ref{SG1} in Eq.(\ref{again}), which proves  Eq.(\ref{qproduct}) and Eq.(\ref{totalj}).  
In conclusion,  the direct product state,  (\ref{prepare}), (\ref{product}),  has a simple classical  interpretation. It corresponds to the classical sum (the vector addition) of two angular momenta, see Fig.~\ref{VSum}.

From the perspective of the general composition rule of two angular momentum states (\ref{composition}), what we have seen are the results concerning a specific pair of states (i.e. those of minimum fluctuations) in ${\underline{j_1}}$ and ${\underline{j_2}}$, characterized by the ``orientations" ${\bf n}_1$ and ${\bf n}_2$. The total angular momentum magnitude in the product state (\ref{qproduct}), (\ref{totalj}) satisfies
\be          
j_1+ j_2 \ge    j_{\rm tot} = |   j_1 {\mathbf n}_1 +    j_2  {\mathbf n}_2|      \ge   |j_1- j_2|\;,  \label{sum}
\ee
 depending on the relative orientation of ${\mathbf n}_1$ and ${\mathbf n}_2$.  Eq.(\ref{sum})  is nicely  consistent with the quantum-mechanical composition  law (\ref{composition}).

 \subsection{Spin-orbit interactions} 
 
 As a possible variation of theme,  and for completeness,  let us consider a spin-orbit interaction of the form,
 \be   H =   A\,  {\mathbf L} \cdot {\mathbf S}\;   \label{SpinOrbit}
 \ee
  where $A$ is a constant,   ${\mathbf L}$ and ${\mathbf S}$ are the orbital angular momentum and spin, respectively,  of a given system. 
  The problem is to understand whether and/or  how such an interaction Hamiltonian reduces effectively to an interaction between two classical angular momenta, in the limit,   $L/\hbar \to \infty$,  $S/\hbar \to \infty$. 
  By writing 
 \be    H =   \frac{A}{2}  \, ({\mathbf J}^2 -    {\mathbf L}^2 -     {\mathbf S}^2 )\;, \qquad  {\mathbf J}=   {\mathbf L} + {\mathbf S}\;,  
 \ee
  and assuming that  ${\mathbf L}^2$ and $ {\mathbf S}^2$ have fixed values $L (L+1) \hbar^2$ and $S (S+1) \hbar^2$
  in the system under consideration,  the problem reduces to that of angular momentum addition, already considered in this section.   
  As such, an interaction of the form (\ref{SpinOrbit}) would reduce to their classical counterpart, when the two states with angular momentum 
  moduli $L$ and $S$ are the states of minimum fluctuations, such as  $| \ell, \ell \ckt $  of (\ref{delta}) and 
   $|j, {\mathbf n}  \ckt $  of  (\ref{Bloch2}). 
   
   On the other hand,  for generic atomic states such conditions are not expected to be satisfied, even in the limit of large atoms. 
Thus the discussion of this section is mostly irrelevant, and the standard, quantum mechanical analysis of spin-orbit interactions are
required always.

\section{Rotation matrix  \label{Rotation}}  

The rotation matrix for a general spin $j$  can be constructed by taking the direct products of the rotation matrix for spin $\frac{1}{2}$, i.e. (\ref{rotationM}). This is  based on the fact that any angular momentum state, $|j, m\ckt$,  can be constructed as the system made of $N$ spin $\tfrac{1}{2}$ particles, completely symmetric under exchange of the particles, a fact already conveniently used in Sec.~\ref{SG1}. 
The rotation matrix for a generic spin $j$  is simply  (writing the Euler angles symbolically as $\Omega \equiv \{\gamma, \theta,\phi\}$) the direct productof the rotation matrices for spin $\frac{1}{2}$, (\ref{rotationM}),  
\be         
(D^{j}(\Omega))_{M, M^{\prime}}   = {\cal S}  \left[  (D^{1/2}(\Omega))_{m_1, m_1^{\prime}}    (D^{1/2}(\Omega))_{m_2, m_2^{\prime}}    \cdots  (D^{1/2}(\Omega))_{m_N, m_N^{\prime}}    \right] 
\ee
symmetrized  (${\cal S}$)  with respect to  $(m_1,m_2,\ldots, m_N)$, where 
\be  
N= 2j \;, \qquad     M=  \sum m_i\;, \qquad  M^{\prime} =  \sum m_i^{\prime}\;. 
\ee
It acts on  the multiplet, 
\bea     
|j,j\ckt   &=&   |\!\uparrow\ckt  |\!\uparrow\ckt  |\!\uparrow\ckt  \cdots   |\!\uparrow\ckt \;,  \nonumber \\
|j,j-1\ckt   &=& \frac{1}{\sqrt N} \{  |\!\downarrow\ckt  |\!\uparrow\ckt  |\!\uparrow\ckt  \cdots   |\!\uparrow\ckt  +   |\!\uparrow\ckt  |\!\downarrow\ckt  |\!\uparrow\ckt  \cdots   |\!\uparrow\ckt +\ldots  \}  \;,  \nonumber \\  
\ldots \nonumber \\
|j, M\ckt   &=&    \frac{1}{\sqrt  {_N C_k }} \{{\rm  sum \,\, of \,\, terms \,\, with}  \,\, k \,\,  |\!\uparrow\ckt's  \}  \qquad   (M=-j+ k)   \nonumber \\ \ldots \nonumber \\
|j,-j\ckt   &=&   |\!\downarrow\ckt  |\!\downarrow\ckt  |\!\downarrow\ckt  \cdots   |\!\downarrow\ckt \;.
\eeq
The action of  $ (D^{j}(\Omega))_{M, M^{\prime}} $  on particular states,  
\be     
|j, {\mathbf n}\ckt  =     |\tfrac{1}{2},  {\mathbf n} \ckt \otimes  |\tfrac{1}{2},  {\mathbf n} \ckt  \otimes \ldots    |\tfrac{1}{2},  {\mathbf n} \ckt\;,  
\ee
where  $ |\tfrac{1}{2},  {\mathbf n}\ckt $ is given in Eq.(\ref{WFspin12}), i.e. on the states with all spin $\tfrac{1}{2}$'s oriented in the same direction, is particularly simple.   It is just the operation of rotating  each and all spins  towards the ${\hat z}$ direction (setting $\gamma=0$),   see (\ref{leads}), 
 \be       
 (D^{j}(\Omega))    |j, {\mathbf n}\ckt     =   |\!\uparrow\ckt  |\!\uparrow\ckt  |\!\uparrow\ckt  \cdots   |\!\uparrow\ckt   =   (1,0,\ldots, 0)^T =   |j, j \ckt  \;.  
 \ee 
This corresponds precisely to the  rotation of a vector,   
\be 
R(\Omega)  \,   {\bf j}^{(\rm class)}   =    R(\Omega)  \,      j\, {\mathbf n}   =  (0,0,j)^T\;,     
\ee    
where $ R(\Omega)  $ is the  standard $3 \times 3$  rotation matrix of classical mechanics.

The discussion of this section is just a simple way of reproducing the well-known fact that the coherent spin (Bloch) states $|j, {\mathbf n}\ckt$  can be obtained by applying the rotation operator  on the state pointing towards ${\hat z}$,   $|j, j_z=j\ckt$  \cite{Radcliffe, Arecchi, Lieb, Puri, Aravind}.


\section{General large-spin states      \label{GenericLS}   }

\begin{figure}
\begin{center}
\includegraphics[width=4.5 in]{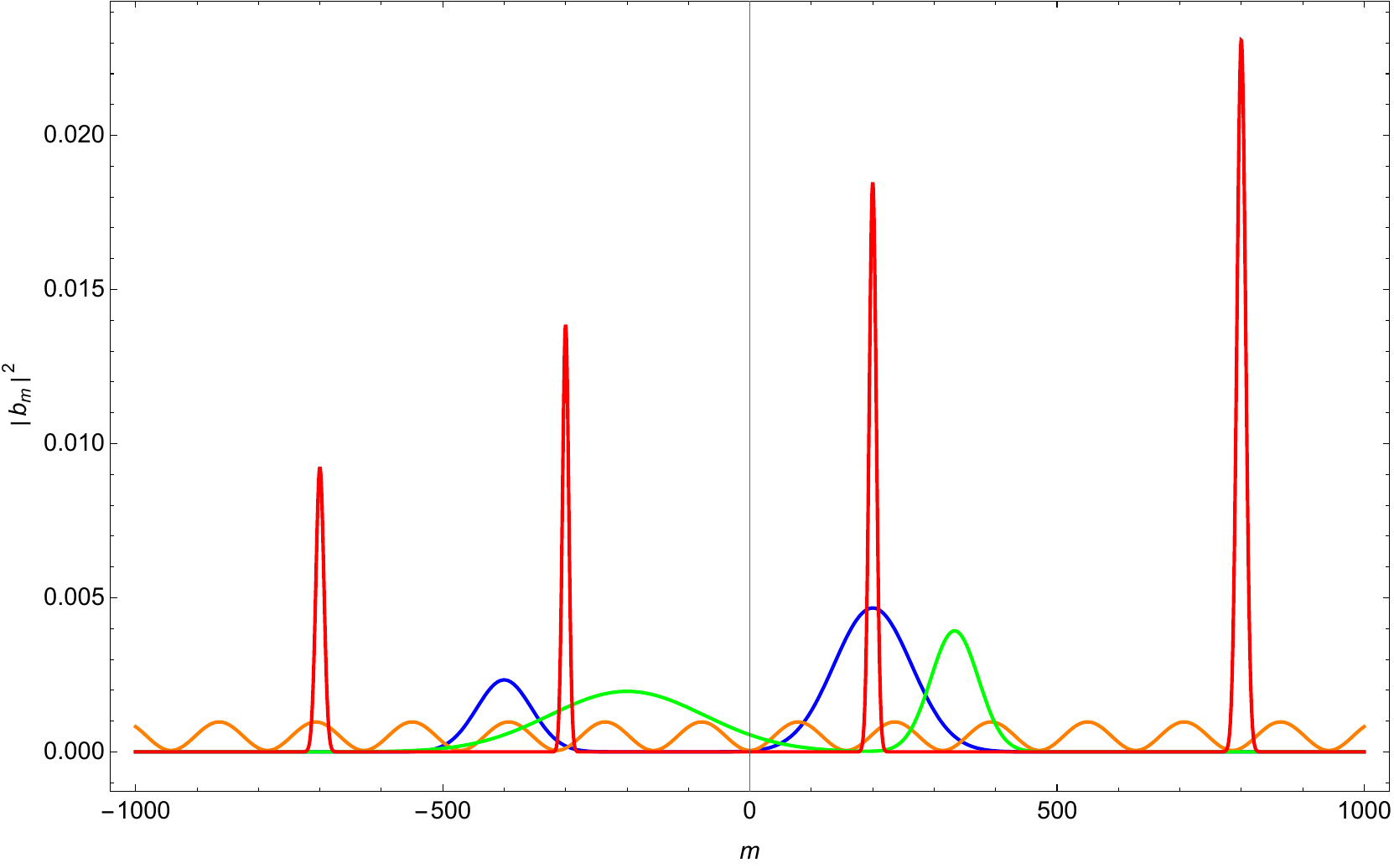}
\caption{\footnotesize {$|b_m|^2$, $J_z=m$,  for a spin $j  =1 \cdot 10^3 $ particle in the state (\ref{genericBis}), with four randomly chosen  sets of coefficients  
$\{b_m\}$.}}
\label{fig:generic}
\end{center}
\end{figure}

In Sec.~\ref{SG} $\sim$   Sec.~\ref{Rotation} we have seen how the particular  spin states, (\ref{proposal}),  behave effectively as a classical angular momentum, 
$j {\mathbf n}$, in the $j\to \infty$  limit.
 It is an entirely different story with a generic large  spin $j$ state in ${\mathbf {CP}}^{2 j}$, 
\be     
| j  \,\, ...  \ckt  = \sum_{m=-j}^{j}       b_m \,   |j,  m\ckt  \;,  \qquad    \sum_{m=-j}^j  |b_m|^2 =1\; .  \label{genericBis} 
\ee
When a particle carrying such a spin enters a Stern-Gerlach set-up with a strong inhomogeneous magnetic field, 
its wave function will in  general  split in many subwavepackets. See Fig.~\ref{fig:generic} for several arbitrarily chosen distributions   $\{b_m\}$, to be 
compared with  Fig.~\ref{Spreadno} or  Fig.~\ref{Spreadno2}.

To discuss the physics of such states,
 it is  useful to recall the so-called Born-Einstein dispute  (see e.g.,  \cite{Joos1}).    Einstein strongly rebuked the idea by Born that the absence of quantum diffusion should be sufficient to explain the classical nature (the unique trajectoriy) of a macroscopic body,  by saying that a doubly or multiply split wave packet, with their centers  separated by a macroscopic distance, are allowed by Schr\"odinger equations,  even for a macroscopic particle. Such a split wave function is certainly non-classical.
 
  The missing piece for solving this apparent puzzle turns out to be the temperature \cite{KK2}.  Even though at exceptionally low temperatures such a state is certainly possible,   it is not so at finite temperatures.   Emission of photons and the ensuing self-decoherence (in the case of an isolated body with finite body temperatures)  \cite{KK2} or an environment-induced decoherence (for an open system) \cite{Joos1,Zurek1,Tegmark},   makes a split system a mixture (a mixed state) essentially instantaneously.   Also, they cannot be prepared experimentally,  e.g.,  by a passage through a double slit \cite{KK2}.  A (macroscopic)  particle passes either through one or the other slit, due to absence of diffusion and/or due to decoherence.
The particle after the passage is behind {\it  either one or the other}   slit, even without any measurement. 

    The status of general angular momentum states with strong fluctuations, (\ref{genericBis}), is similar to that of Einstein's 
    macroscopic split wave packet.  In the large $j$ limit, a generic large spin state  (represented by randomly distributed stars, Fig.~\ref{Stars}),  far from the Bloch state  $|j, \mathbf n\ckt$,      necessarily acquires a large space extension under an inhomogeneous magnetic field: a macroscopic quantum state.  See Fig.~\ref{fig:generic}.   But 
a  macroscopic pure quantum state is possible only at exceedingly low temperature close to $T=0$: otherwise,  it is a mixture.

To show quantitatively how a macroscopically split wave packet becomes a mixed state  under an environment-induced decoherence, is 
a complex problem, depending on the details of the environment itself, the temperature, density, pressure, kinds of environment particles, and average density and momentum distribution, the mean de Broglie wave length, etc.  It is beyond the scope of the present work to perform such a study.  However,  a qualitative discussion about how a macroscopically split wave packet becomes a mixture
under an environment-induced decoherence  \cite{Joos1,Zurek1,Tegmark}, but not classical,    has been given in \cite{KKHTE},  in the case of a spin $\frac{1}{2}$  particle.

To tie up possible loose ends of the discussion,  note that  a single, spin $\frac{1}{2}$  isolated atom in its ground state 
(a microscopic system) travelling 
in a good vacuum, is effectively a system at  $T=0$. Thus the fact that its split wave packets can get separated by a macroscopic distance in an SG process - i.e. its being in a macroscopic quantum state - is perfectly consistent with the general argument above. To realize a macroscopic system composed of many atoms and molecules  in a pure quantum state, is another story:  it is much more difficult to prepare the necessary low temperatures close to $T=0$,  to maintain its pure-state nature.
At finite temperatures, a macroscopically split wave packet of such  a particle, which might arise  under the SG  set-up,  as in Fig.~\ref{fig:generic},   is necessarily an incoherent mixture. 

An important point to keep in mind, however, is the following.  The fact that a macroscopic quantum state such as the split wavepackets of Fig.~\ref{fig:generic}  becomes a mixture under the decoherence effects at finite temperatures  \cite{Joos1,Zurek1,Tegmark}, does not mean that it becomes classical. Decoherence and classical limits are two distinct phenomena in general  \cite{KKHTE}.  This point can hardly be overemphasized. 
    
What is indeed remarkable, perhaps, is the fact that the angular momentum states of minimum fluctuations,  (\ref{Bloch}), (\ref{Bloch2}), (\ref{see00}), (\ref{exercise00}),  though pure, effectively become classical at $j\to \infty$,  without help of any (thermal or environment-induced) decoherence effects, hence independently of the temperatures.  This might appear to be in line with the familiar discussion on the semi-classical limit of a (pure-state) wave function in QM at  $\hbar \to 0$, but,  our discussion about the generic large spin states indicates the presence of a loophole in the argument that 
 only the dimensionless ratio  $j/\hbar$ should matter.

\section{Conclusion \label{Conclusion} } 

The observations made in this work render  the concept that a spin  (or  angular momentum), $j$,   becomes classical in the limit, $j  \to \infty$, a more precise one.  In particular, we found that the states with minimum fluctuations  (\ref{Bloch2}) and the states close to them,  are to be identified  with classical angular momentum vectors in such a limit, and verified its consistency through the analyses of the SG experiments, with the angular-momentum composition rule, and with the rotation matrix.   

 At the same time,  our analysis has revealed a subtlety in the quantum-classical correspondence in the large spin limit in general.  It  reflects  the 
difference  between the spaces of the quantum-mechanical  and classical angular momentum states of definite magnitude    ($\mathbf{CP}^{2j}$  and  $\mathbf{CP}^1\sim S^2$, respectively).
It is the angular momentum  states of minimum fluctuations, (\ref{Bloch}),(\ref{Bloch2}),(\ref{see00}),(\ref{exercise00}),   that naturally smoothly become  classical angular momentum in the   $j \to \infty$  limit, independently of  the temperatures.  

 On the other hand,  more general, strongly fluctuating angular momentum states belonging to $\mathbf{CP}^{2j}$ remain quantum-mechanical even in the $j \to \infty$  limit.   They can stay pure quantum states only at exceedingly low temperatures, and isolated.  
 The reason is that in the  $j\to \infty$ limit such a state necessarily involves a large space extension, in a way or other.  A macroscopic system is  
 in a mixed state at finite temperatures \cite{Joos1,Zurek1,Tegmark,Joos},\cite{KK2}.  No macroscopically split pure wave packets in a SG setting survive decoherence.  At the same time, as pointed out in  \cite{KKHTE},  a macroscopic quantum state which becomes  a mixture
 due to decoherence, remains such:  a mixed {\it  quantum} state.   Decoherence in itself  does not render the system involved classical, contrary to such an idea sometimes expressed in the literature \cite{Zurek1,Joos}. 

The idea that a large spin is made of many spin $\tfrac{1}{2}$ particles turned out to be  quite a useful one both mathematically and physically, as we saw in several occasions.   The entire theory of angular momentum can indeed be reconstructed this way as shown by Schwinger   \cite{Schwinger},  and it helps to understand certain formulas easily.  It is also relevant  from the physics point of view,  as such a large angular momentum 
system may be regarded as an idealized toy-model version of a macroscopic body, made of many atoms and molecules (spins).   Seen this way,  the sharp spike of the SG projection of the state $|j, {\bf n}\ckt $  in the narrow region 
around $J_z=  j \cos \theta $   we have observed,  such as in   Fig.~\ref{Spreadno},  Fig.~\ref{Spreadno2},    can  be interpreted as a results of a large number of microscopic states of atoms (spins)
 accumulating to give such a value of  $J_z$. \footnote{The  suppression of the  fluctuations of angular momentum  components we observed in this note  in the $j\to \infty$ limit,    is somewhat analogous to the ``$\sqrt{n}$ law"  invoked by Schr\"odinger \cite{Life},  to explain the general exactness (classicality) of laws of macroscopic world based on statistical mechanics,  although, here, the relevant fluctuations are quantum-mechanical ones. 
This being so,  let us not forget that the deepest insight of Schr\"odinger in his book was that many  biological processes such as mutation, reproduction,  and  inheritance, involve quantum mechanics  in essential ways. }

Nevertheless, our discussion is really about the quantum and classical properties {\it  of a single large spin,}  and do not concern the thermodynamical, statistical or other physical  properties of realistic many-body systems, as in \cite{Arecchi,Lieb}.
And in spite of inevitable partial overlaps with numerous  observations made in the literature  \cite{Brussard}-\cite{Corso},  our careful observations about how the particular class of  large momentum (spin) states get classical in the limit, $j  \to \infty$, 
independently of the temperature, whereas a generic large $j$ states remain quantum mechanical in the same limit,  as
illustrated in Sec.~\ref{spaces} - Sec.~\ref{Addition}, and  Sec.~\ref{GenericLS} are, to the best of our knowledge, new.

\section*{Acknowledgments} 
 The work by K.K. is supported by the INFN special-initiative project grant GAST (Gauge and String Theories).  We are grateful to Hans Thomas Elze, Jarah Evslin, Riccardo Guida, Pietro Menotti and Arkady Vainshtein for useful comments and precious advices.  

%

\appendix
\section {Newton's equation for a macroscopic body  \label{NewtonEq}}  
The conditions needed for the CM of an isolated macroscopic body {\it  at finite body temperatures} to obey   
Newton's equations have been investigated in great care by one of the present authors  \cite{KK2}. They are 
\begin{itemize}
  \item[(i)] For macroscopic motions  (for which $\hbar \simeq 0$)  the Heisenberg relation does not limit the simultaneous determination -- the initial condition -- of the position  and momentum;
  \item[(ii)]   The absence of quantum diffusion, due to a large mass (a large number of atoms and molecules composing the body);  
  \item[(iii)]  A finite body temperature, implying the thermal decoherence and mixed-state nature of the body. 
\end{itemize}
Under these conditions, the CM of an isolated macroscopic body has a unique trajectory. 
Newton's equations for it  follow from the Ehrenfest theorem. See Ref.~\cite{KK2} for various subtleties and  for the explicit derivation of Newton's equation
under external gravitational forces,  under weak, static, smoothly varying external electromagnetic fields, and under a harmonic-oscillator potential.
Somewhat unexpectedly, the environment-induced decoherence \cite{Joos1,Zurek1,Tegmark}  which is extremely effective in rendering macroscopic states in a finite-temperature environment  a mixture, is found not to be the most essential element for the derivation of classical mechanics  from quantum mechanics \cite{KK2, KKHTE}.  

\section{A subtle face of the Stern-Gerlach experiment   \label{SGB}} 

 The Hamiltonian is given by
 \be   H=   \frac{{\bf p}^2}{2m}  + V\;, \qquad        V=  -  {\boldsymbol  {\mu}  \cdot {\bf B}}\;,\label{Hamiltonian}
\ee
\be    {\boldsymbol  \mu} =   \mu_B    \, g \,  {\mathbf s}\;,     \qquad \partial  B_z / \partial z \ne 0\;,  \label{Spin}
\ee
where $ \mu_{B} =  \frac{ e \hbar }{2 m_e c}$ is the Bohr magneton.  We recall the 
well-known fact that the gyromagnetic ratio $g \simeq 2$ of the electron and the spin magnitude $1/2$ approximately cancel,
so $\mu_B$  is the magnetic moment \footnote{For illustration purpose, we use this value below, but the discussion is obviously valid for general atoms.}, in the case of the atoms such as $A_g$,  where a single outmost electron provides the total spin $\tfrac{1}{2}$. 

An example of the  inhomogeneous field ${\bf B}$ appropriate for the Stern-Gerlach experiment is  \cite{Platt,Alstrom} 
\be      {\bf B}  =   (  0,   B_y,  B_z ), \qquad   B_y = - b_0 \, y, \quad B_z = B_0 +  b_0  \, z\;     \label{magneticfield}
\ee   
which satisfy   $\nabla \cdot  {\bf B} =  \nabla \times {\bf B} =0$.  The constant field $B_0$ in the $z$ direction must be large, 
\be   |B_0|  \gg       |b_0 \, y|\;.  \label{largeB}
\ee
 in the relevant region of $(y,z)$ of the experiment.  
The wave function of the spin $\frac{1}{2}$ particle entering the SG magnet  has the form,   
\be    
  \Psi=  {\tilde \psi}_1({\bf r}, t) |\!\uparrow\ckt +   \, {\tilde \psi}_2({\bf r}, t)   |\!\downarrow\ckt \;.
\ee
obeying the Schr\"odinger equation,
\be   i \hbar  \frac{d}{dt}  \Psi =   H\, \Psi\;. 
\ee
By redefining the wave functions for  the upper and down  spin components  as
\be      {\tilde \psi}_1({\bf r}, t)  =   e^{i  \mu_B  B_0 t / \hbar}   \psi_1({\bf r}, t) \;, \qquad   {\tilde \psi}_2({\bf r}, t)  =   e^{-  i  \mu_B  B_0 t / \hbar}   \psi_2({\bf r}, t) \;,
\ee
  one finds that the up- and down- spin components  $\psi_{1}$ and  $\psi_{2}$  satisfy the separate  Schr\"odinger equations \cite{Platt} 
\be   i \hbar  \frac{\de}{\de t}   \psi_1 =     \left(    \frac{{\bf p}^2}{2m}   -      \mu_{B}  b_0    z    \right)       \psi_1 \;,\qquad  
  i \hbar  \frac{\de}{\de t}   \psi_2  =     \left(    \frac{{\bf p}^2}{2m}   +       \mu_{B}  b_0    z    \right)       \psi_2 \;.
     \label{SEq}
\ee
This is because 
the term   $\propto  -\mu_y B_y=- (g \mu_B  b)  s_y $  in the Hamiltonian (\ref{Hamiltonian})    
mixing the two components $\psi_{1,2}$  
  has acquired a rapidly oscillating phase factor, 
\be      \pm i  \mu_B\, b_0  y \,   e^{\mp   2  i  \mu_B  B_0 t / \hbar}   \;,
\ee
hence can be safely neglected.    
  The condition (\ref{largeB}) is crucial here.

  As explained in \cite{Alstrom},  this can be  classically understood as the spin precession  effect around the large constant magnetic field $B_0 \hat z$, thanks to which  the forces on the particle in the transverse ($\hat x, \hat y$) directions average out to zero  \footnote{
With a magnetic field $B_0$ of the order of $10^3$ Gauss, 
the precession frequency is of the order of $10^{11}$ in the case of the SG experiment with silver atom \cite{Alstrom}.  With the average velocity of $A_g$ atoms of the order of  $100 \, m/s$  and the size of the region of the magnetic field of the order of a few $cm$ \cite{SG},  the 
timescale of the precession is orders of magnitude ($\sim 10^{-5}$)  shorter than the time the atoms spend in the region.}. The only significant force  it receives is due to the inhomogeneity in  $B_z$, (\ref{Spin}), which deflects the atom in the $\pm \hat{z}$ direction.

From (\ref{SEq}) and their complex conjugates,  one can derive the Ehrenfest theorem for $\psi_1$ and $\psi_2$ separately,  
      \bea  &&   \frac{d}{dt}   \brc {\bf r}\ckt_1   =  \brc {\bf p}/{m} \ckt_1  \;,   \quad   
      \frac{d}{dt}  \brc {\bf p}\ckt_1  =   -    \brc  \nabla (\mu_{B}   B_z) \ckt_1  \;;      \label{New1}   \\
   &&    \frac{d}{dt}   \brc {\bf r}\ckt_2   =  \brc {\bf p}/{m} \ckt_2  \;,   \quad   
      \frac{d}{dt}  \brc {\bf p}\ckt_2  =   +  \brc  \nabla (\mu_{B}   B_z) \ckt_2  \;,   \label{New2}
\eea
where    $\brc {\bf r}\ckt_1  \equiv   \brc   \psi_1 | {\bf r} | \psi_1 \ckt$, etc.    
Namely, for a sufficiently compact initial wave packet,  $\psi({\bf r}, t)$, the expectation values of    ${\bf r}$ and ${\bf p}$ in
the up and down components   $\psi({\bf r}, t)_{1,2}$  trace  respectively  the classical trajectories of a spin-up or spin-down particle.  

 Still,  even though the two subpackets might get well separated by a macroscopic distance, $\Psi$ remains  in a coherent superposition of the upper and lower spin components  $\psi_{1,2}$. Its pure-state nature can be verified, in a set-up known as the quantum eraser,    
   by reconverging them by using  a second magnetic field of opposite gradients, and studying their interferences   \cite{Kwiat}.

A variational  solution of (\ref{SEq}) for Gaussian split wave packets can be found in  \cite{KKHTE}.   


\begin{thebibliography}{22}

\bibitem{Brussard}
P.~J. Brussard and H.~A. Tolhoek, 
``Classical limits of Clebsch-Gordan coefficients,  Racah coefficients and  
$D^{l}_{m n}(\varphi, \vartheta, \psi)$-functions'',
\href{https://doi.org/10.1016/S0031-8914(57)95547-7}{Physica~{\bf 23}, 955 (1957)}.

\bibitem{Barut}
A.~O. Barut and J.~Dilley,
``Behavior of the Scattering Amplitude for Large Angular Momentum'', 
\href{https://doi.org/10.1063/1.1703920}{J. Math. Phys.~{\bf 4}, 1401 (1963)}.

\bibitem{CederRamsey}
J.~W. Cederberg and N.~F. Ramsey, 
``Magnetic Resonance with Large Angular Momentum'', 
\href{https://doi.org/10.1103/PhysRev.135.A39}{Phys. Rev.~{\bf 135}, A39 (1964)}.

\bibitem{Newton}
R.~G. Newton and B.~-lin Young,
``Measurability of the spin density matrix'',
\href{https://doi.org/10.1016/0003-4916(68)90035-3}{Annals of Physics, {\bf 49} 393 (1968)}. 

\bibitem{Radcliffe}
J.~M. Radcliffe, 
``Some properties of coherent spin states'', \href{https://doi.org/10.1088/0305-4470/4/3/009}{J.~Phys.~A: Gen. Phys.~{\bf  4} (1971)}. 

\bibitem{Arecchi} 
F.~T. Arecchi, E. Courtens, R. Gilmore and H. Thomas,
``Atomic Coherence States in Quantum Optics'',  \href{https://doi.org/10.1103/PhysRevA.6.2211}{Phys. Rev. A~{\bf 6}, 2211 (1972)}. 

\bibitem{Lieb}
E.~H. Lieb,  
``The Classical Limit of Quantum Spin Systems'', \href{https://doi.org/10.1007/BF01646493}{
Commun. Math. Phys.~{\bf 31}, 327 (1973)}.  

\bibitem{Wu:1982fyj}
W.~Hua-Chuan,
``Bosons with large angular momentum and rotation of even-even nuclei'',
\href{https://doi.org/10.1016/0370-2693(82)90938-8}{Phys. Lett. B~\textbf{110}, 1-6 (1982)}.

\bibitem{Wodk}
K.~Wodkiewicz and J.~H.~Eberly,
``Coherent states, squeezed fluctuations, and the SU(2) am SU(1,1) groups in quantum-optics applications",
\href{https://doi.org/10.1364/JOSAB.2.000458}{J. Opt. Soc. Am. B \textbf{2}, 458-466 (1985)}.

\bibitem{Allen:1992zz}
L.~Allen, M.~W. Beijersbergen, R.~J.~C. Spreeuw and J.~P. Woerdman,
``Orbital angular momentum of light and the transformation of Laguerre-Gaussian laser modes'',
\href{https://doi.org/10.1103/PhysRevA.45.8185}{Phys. Rev. A~\textbf{45}, 8185 (1992)}.

\bibitem{Puri}
R.~R. Puri, ``Coherent and squeezed states on physical basis'', \href{https://doi.org/10.1007/BF02845612}{Pramana - J Phys~{\bf 48} 787 (1997)}.


\bibitem{Aravind}
P.~K. Aravind,
``Spin coherent states as anticipators of the geometric phase'', \href{https://doi.org/10.1119/1.19145}{Am. J. Phys.~{\bf 67}, 899 (1999)}. 

\bibitem{Klose}
G.~Klose, G.~Smith and P.~S.~Jessen,
``Measuring the Quantum State of a Large Angular Momentum'',
\href{https://doi.org/10.48550/arXiv.quant-ph/0101017}{Phys. Rev. Lett.~{\bf 86}, 4721 (2001)}; [arXiv:quant-ph/0101017 [quant-ph]].

\bibitem{Imamura:2002xq}
Y.~Imamura,
``Large angular momentum closed strings colliding with D-branes'', \href{https://10.1088/1126-6708/2002/06/005}{
JHEP \textbf{06}, 005 (2002)}; [arXiv:hep-th/0204200 [hep-th]].

\bibitem{Livine}
E.~R.~Livine and S.~Speziale,
``New spinfoam vertex for quantum gravity'', \href{https://doi.org/10.1103/PhysRevD.76.084028}{Phys. Rev. D \textbf{76}, 084028 (2007)}; [arXiv:0705.0674 [gr-qc]].


\bibitem{Anderson}
R.~W. Anderson, V.~Aquilanti and  C.~da~Silva~Ferreira,
``Exact computation and large angular momentum
asymptotics of $3nj$ symbols: Semiclassical disentangling of
spin networks'', \href{
https://doi.org/10.1063/1.3000578
}{J. of Chem. Phys.~{\bf 129}, 161101 (2008)}.    

\bibitem{Dajka}
J.~Dajka,
``Disentanglement of Qubits in Classical Limit
of Interaction'', \href{https://doi.org/10.1007/s10773-013-1876-9}{
Int. J. Theor Phys~{\bf 53} 870 (2014)}.

\bibitem{Kovacs:2013una}
S.~Kovacs, Y.~Sato and H.~Shimada,
``Membranes from monopole operators in ABJM theory: Large angular momentum and M-theoretic $AdS_4/CFT_3$'', \href{https://doi.org/10.1093/ptep/ptu102}{Progress of Theoretical and Experimental Physics, Issue 9 093B01 (2014)}; [arXiv:1310.0016 [hep-th]].

\bibitem{LohKim}
Y.~L. Loh and M. Kim,
``Visualizing spin states using the spin coherent state
representation'', \href{https://doi.org/10.1119/1.4898595}{Am. J. Phys.~{\bf 83}, 30 (2015)}.

\bibitem{Byrnes}
T.~Byrnes, D.~Rosseau, M.~Khosla, A.~Pyrkov, A.~Thomasen, T.~Mukai, S.~Koyama, A.~Abdelrahman and E.~Ilo-Okeke,
``Macroscopic quantum information processing using spin coherent states,''
\href{https://doi.org/10.1016/j.optcom.2014.08.017}{Opt. Commun. \textbf{337}, 102-109 (2015)}


\bibitem{Chen:2018tkb}
Y.~Y. Chen, J.~X. Li, K.~Z. Hatsagortsyan and C.~H. Keitel,
``\ensuremath{\gamma}-Ray Beams with Large Orbital Angular Momentum via Nonlinear Compton Scattering with Radiation Reaction'', \href{https://doi.org/10.1103/PhysRevLett.121.074801}
{Phys. Rev. Lett.~\textbf{121}, 074801 (2018)}; [arXiv:1802.04748 [physics.plasm-ph]].


\bibitem{KusMostow}
M. Kus, J. Mostowski and J. Pietraszewicz, 
``Classical limit of entangled states of two angular momenta'', \href{https://doi.org/10.1103/PhysRevA.99.052112}{Phys. Rev. A~{\bf 99}, 052112 (2019)}.  


\bibitem{Braccini:2023eyc}
L. Braccini, M. Schut, A. Serafini, A. Mazumdar and S. Bose,
``Large Spin Stern-Gerlach Interferometry for Gravitational Entanglement'', \href{https://doi.org/10.48550/arXiv.2312.05170}{arXiv:2312.05170 [quant-ph] (2023)}.

\bibitem{Corso}
P.~P. Corso, D. Cricchio and E. Fiordilino,
``Large Angular Momentum States in a Graphene Film'',
\href{https://doi.org/10.3390/physics6010021}{Physics~{\bf  6}, 317 (2024)}.


\bibitem{KK2}
K. Konishi, ``Newton's equations  from
quantum mechanics  for a macroscopic body in the vacuum", \href{https://doi.org/10.1142/S0217751X2350080X}{Int. Journ. Mod. Phys. A~{\bf  38}  2350080 (2023)}; [arXiv:2209.07318 [quant-ph] (2022)].

\bibitem{KKHTE}
K. Konishi and H.~T. Elze, ``The Quantum Ratio'', \href{https://doi.org/10.3390/sym16040427}{Symmetry~{\bf 2024}, 16(4), 427};
{[arXiv:2402.10702 [quant-ph] (2024)]}.

\bibitem{Bengt}
I. Bengtsson and K. Zyczkowski
``Geometry of quantum states: An introduction to quantum entanglement,'' \href{https://doi.org/10.1017/CBO9780511535048}{Cambridge University Press (2006)}.

\bibitem{Dubrovin}
 B.~A. Dubrovin, S.~P. Novikov and A.~T. Fomenko, 
 ``Modern Geometry -- Methods and Applications. Part II: The Geometry and Topology of Manifolds'',
 \href{https://doi.org/10.1007/978-1-4612-1100-6}{Graduate Texts in Mathematics, Springer New York NY (1985)}.


\bibitem{Schwinger}  
J. Schwinger,  ``On Angular Momentum", World Scientific Series in 20th Century Physics, \href{https://doi.org/10.1142/9789812795694_0010}{A Quantum Legacy, 173-223 (2000)}.
A reproduction of a book chapter from
L.~C.~Biedenharn,  H.~van Dam,  ``Angular Momentum in Quantum Physics",  
\href{https://doi.org/10.2172/4389568}
{New York,  Academic Press (1965)}. 



\bibitem{SG}
W. Gerlach and O. Stern,  ``Der experimentelle Nachweis der Richtungsquantelung im Magnetfeld", \href{https://doi.org/10.1007/BF01326983}{Zeitschrift für Physik.~{\bf 9} 349 (1922)}.

\bibitem{Alstrom}
P. Alstr{\o}m, P. Hjorth and R. Mattuck, 
 ``Paradox in the classical  treatment of the Stern-Gerlach experiment'', 
 \href{http://dx.doi.org/10.1119/1.12732}{Am. J. Phys.~{\bf 50}, 697 (1982)}.
 
\bibitem{Platt}
D.~E. Platt, 
``A modern analysis of the Stern-Gerlach experiment'',
\href{http://dx.doi.org/10.1119/1.17136}{Am. J. Phys.~{\bf 60}, 306 (1992)}.

\bibitem{KK3}
K. Konishi, ``On the negative-result experiments in quantum mechanics'', \href{https://doi.org/10.3390/e26110958}{Entropy~{\bf 2024}, 26(11), 958}; [arXiv:2310.01955 [quant-ph]] (2023)].
%
%
\bibitem{Joos1} 
E. Joos and  H.~D. Zeh, ``The emergence of classical properties through interaction with the environment", \href{https://doi.org/10.1007/BF01725541}{Z. Phys. B~{\bf 59}, 223 (1985)}.
%
\bibitem{Zurek1}
W.~H. Zurek, ``Decoherence and the Transition from Quantum to Classical", \href{https://doi.org/10.1063/1.881293}{Physics Today~{\bf 44} (10), 36 (1991)}.
%
\bibitem{Tegmark}  
M. Tegmark, ``Apparent wave function collapse caused by scattering'', \href{https://doi.org/10.1007/BF00662807}{Found. Phys. Lett.~{\bf 6}, 571 (1993)}.


\bibitem{Joos}  
E.~Joos, H.~D.~Zeh, C.~Kiefer, D.~Giulini, J.~Kupsch, I.~O.~Stamatescu, ``Decoherence and the Appearance of a Classical World in Quantum Theory", \href{https://doi.org/10.1007/978-3-662-05328-7}{Springer (2003)}. 


\bibitem{Life}
E. Schr\"odinger, ``What is life?", \href{https://doi.org/10.1017/CBO9781107295629}{Cambridge University Press, (1944)}.

\bibitem{Kwiat}
P.~G. Kwiat, A.~M. Steinberg and R.~Y. Chiao,  
``Observation of a ``quantum eraser": A revival of coherence in a two-photon interference experiment'', \href{https://doi.org/10.1103/PhysRevA.45.7729}{Phys. Rev. A~{\bf 45},  7729 (1992)}.

\end{thebibliography}
\end {document}